\documentclass[preprint]{aastex}
\usepackage{natbib,amssymb}
\usepackage{amsmath}
\shorttitle{SONYC VIII: Lupus$\,$3}
\shortauthors{Mu\v{z}i\'c et al.}

\begin{document}
\bibliographystyle{apj}

\newcommand{\solm}{M$_{\odot}$}
\newcommand{\soll}{L$_{\odot}$}
\newcommand{\solr}{R$_{\odot}$}
%================================================================

\title{Substellar Objects in Nearby Young Clusters (SONYC) VIII: Substellar population in Lupus 3\altaffilmark{*}
}

\author{Koraljka Mu\v{z}i\'c\altaffilmark{1}, Alexander Scholz\altaffilmark{2,3}, Vincent C. Geers\altaffilmark{2}, Ray Jayawardhana\altaffilmark{4}, Bel\'en L\'opez Mart\'i\altaffilmark{5,6}}

\email{kmuzic@eso.org}

\altaffiltext{1}{European Southern Observatory, Alonso de C\'ordova 3107, Casilla 19, Santiago, 19001, Chile}
\altaffiltext{2}{School of Cosmic Physics, Dublin Institute for Advanced Studies, 31 Fitzwilliam Place, Dublin 2, Ireland}
\altaffiltext{3}{School of Physics \& Astronomy, St. Andrews University, North Haugh, St Andrews KY16 9SS, United Kingdom}
\altaffiltext{4}{Department of Astronomy \& Astrophysics, University of Toronto, 50 St. George Street, Toronto, ON M5S 3H4, Canada}
\altaffiltext{5}{Centro de Astrobiolog\'ia (INTA-CSIC), Departamento de Astrof\'isica, PO Box 78, 28261 Villanueva de la Ca\~nada, Madrid, Spain}
\altaffiltext{6}{Saint Louis University -- Madrid Campus, Division of Science, Engineering and Nursing, Avenida del Valle 34, E-28003 Madrid, Spain}
\altaffiltext{*}{Based on observations collected at the European Southern Observatory under programs
087.C-0386 and 089.C-0432, and Cerro Tololo Interamerican Observatory's programmes 2010A-0054 and 2011A-0144.}

\begin{abstract}
SONYC -- Substellar Objects in Nearby Young Clusters -- is a survey programme to investigate the frequency 
and properties of substellar objects in nearby star-forming regions.
We present a new imaging and spectroscopic survey conducted in the young ($\sim1\,$Myr), nearby ($\sim200\,$pc) 
star-forming region Lupus$\,$3.
Deep optical and near-infrared images were obtained with MOSAIC-II and NEWFIRM
at the CTIO-4m telescope, covering $\sim1.4\,$deg$^2$ on the sky. The $i$-band completeness limit of 20.3$\,$mag is
equivalent to $0.009-0.02$ \solm, for A$_V\leq5$. 
Photometry and $11-12\,$yr baseline proper motions were used to select candidate
low-mass members of Lupus~3.
We performed spectroscopic follow-up of 123 candidates,
using VIMOS at the Very Large Telescope (VLT), and
identify 7 probable members, 
among which 4 have spectral type later than M6.0 and $T_{\mathrm{eff}}\leq3000\,$K, 
i.e. are probably substellar in nature. 
Two of the new probable members of Lupus$\,$3 appear underluminous for their 
spectral class and exhibit emission line spectrum with strong H$_{\alpha}$ or forbidden lines 
associated with active accretion.  
We derive a relation between the spectral type and effective temperature: T$_{\mathrm{eff}}=(4120\pm175)-(172\pm26)\times\mathrm{SpT}$, 
where SpT refers to
the M spectral subtype between 1 and 9.
Combining our results with the previous works on Lupus$\,$3,
we show that the spectral type distribution is consistent with that in other star forming regions, as well as is the 
derived star-to-BD ratio of $2.0-3.3$. 
We compile a census of all spectroscopically confirmed low-mass members with spectral type M0 or later.
\end{abstract}
%Photometric candidate selection was combined with the proper motions based on a $11-12\,$yr
%baseline.

%===================================================================

\section{Introduction}
\label{intro}

SONYC - short for {\it Substellar Objects in Nearby Young Clusters} - is a comprehensive project aiming
to provide a complete, unbiased census of substellar population down to a few Jupiter masses
in young star forming regions. Studies of the substellar mass regime at young ages are crucial to understand
the mass dependence in the formation and early evolution of stars and planets. 
Although the low-mass end of the Initial Mass Function (IMF) has been the subject of intensive investigation
over more than a decade, and by various groups, its origin is still a matter of debate 
(e.g. \citealt{bonnell07, bastian10, jeffries12b}). The relative importance of several proposed processes 
(dynamical interactions, fragmentation, accretion, photoevaporation) 
responsible for the formation of brown dwarfs (BDs) is not yet clear.

The SONYC survey is based on extremely deep optical- and near-infrared wide-field imaging, combined
with the Two Micron All Sky Survey (2MASS) and $Spitzer$ photometry catalogs, which are correlated to create 
catalogs of substellar candidates and used to identify targets for extensive spectroscopic follow-up. 
In this work for the first time we also include a proper motion analysis, which greatly facilitates the 
candidate selection. Our observations are designed to reach mass limits well below 0.01$\,$M$_{\odot}$, and the main 
candidate selection method is based on the optical photometry. This way we ensure to obtain a realistic picture of the 
substellar population in each of the studied cluster, avoiding the biases introduced by the mid-infrared selection 
(only objects with disks), or methane-imaging (only T-dwarfs). 
So far we have published results for three regions: 
NGC 1333 \citep{scholz09,scholz12a,scholz12b}, $\rho$ Ophiuchi \citep{geers11,muzic12}, and Chamaeleon-I \citep{muzic11}.
We have identified and characterized more than 50 new substellar objects, among them a handful
of objects with masses close to, or below the Deuterium burning limit.
Thanks to the SONYC survey and the efforts of other groups, the substellar IMF is now well characterized 
down to $\sim5 - 10 $M$_J$, and we find that the ratio of the number of stars with respect to brown dwarfs lies between 2 and 6.  
In NGC1333 we find that, down to $\sim5$M$_J$, the free-floating objects with planetary masses

In this paper we present the SONYC campaign in the Lupus$\,$3 star forming region.
The Lupus dark cloud complex is located in the Scorpius-Centaurus
OB association and consists of several loosely connected dark clouds showing
different levels of star-formation activity (see \citealt{comeron08} for a detailed overview).
%low-mas pre-main sequence stars \citep[e.g.]{comeron03, comeron08, lopezmarti05, merin08, comeron09}.
The main site of star formation within the complex is Lupus$\,$3, which contains one of the richest 
associations of T-Tauri stars \citep{schwartz77, krautter97}. The center of Lupus$\,$3 is 
dominated by the two most massive members of the entire complex, the two Herbig Ae/Be stars known as 
HR~5999 and HR~6000. More than half of the known Lupus$\,$3 members are found in the $0.3 \times 0.3$ pc$^2$ area
surrounding the pair \citep{nakajima00, comeron08}. 

The low-mass (sub-)stellar content of Lupus$\,$3 was extensively investigated using the data from 
the Spitzer Space Telescope. % \citep{allers06, chapman07, allen07,merin08}. 
\citet{merin08} compiled the most complete census of stars and brown dwarfs in Lupus$\,$3 at the time, 
using the data from their survey, together with the results of previous surveys by \citet{nakajima00, comeron03,comeron08,lopez-marti05, allers06,allen07,chapman07,tachihara07,merin07,strauss92,gondoin06}. Spectroscopic follow-up in the optical (FLAMES/VLT) 
by \citet{mortier11} confirmed the effectiveness of
MIR-excess selection from \citet{merin08}, with about 80\% 
of the 46 observed sources confirmed as members of Lupus$\,$3.
Two wide-area photometric surveys
in Lupus were conducted using the Wide Field Imager (WFI) at the La Silla 2.2-m telescope.
\citet{lopez-marti05} identified 22 new low-mass member candidates in an area of 1.6 deg$^2$ in Lupus$\,$3, with about half of the candidates 
confirmed spectroscopically in surveys by \citet{allen07} and \citet{mortier11}.   
\citet{comeron09} surveyed an area of more than 6 deg$^2$ in the Lupus~1, 3, and 4 clouds
and identified $\sim$70 new candidate members of Lupus$\,$3. About 50\% of the
photometric sample was revealed to belong to a background population of giant stars in a follow-up spectroscopic study by \citet{comeron13}.  
The deepest
survey so far in Lupus$\,$3 \citep{comeron11}, did not report any brown dwarfs, despite the sensitivity to objects with sub-Jupiter masses. 
However, the surveyed area of $\sim 7' \times 7'$ was very small compared to the total extent of the cluster, and
thus contained only a minor fraction of the total population. The substellar population in Lupus$\,$3 remains therefore poorly constrained.
%Clearly, all previous studies in Lupus$\,$3 are either biased towards sources with disk excess \citep{merin08, mortier11}, and H$_{\alpha}$ emission
%\citep{lopez-marti05}, or lack spectroscopic confirmation of membership \citep{comeron09}.

The distance to the Lupus star forming region is still a matter of debate, with the distance determinations from different studies
ranging from 140 pc to 240 pc \citep{comeron08}. A widely-used value of $140\pm20\,$ pc was
derived by \citet{hughes93} using photometry and spectroscopy of F-G field stars in Lupus. Using an improved version of the same
method, \citet{lombardi08} obtained a distance of $155\pm8\,$ pc, but also suggest that the apparent
thickness of Lupus might be the result of different Lupus sub-clouds being at different distances. 
Distances obtained from the moving-cluster method \citep{makarov07} suggest a thickness of $\sim80\,$pc, and places
 Lupus$\,$3 at a distance of about 170 pc, or 25 pc farther than the center of the greater association. 
This spatial arrangement is in general agreement with the distances derived from Hipparcos parallaxes of individual stars located
in different sub-clouds \citep{wichmann98, bertout99}. 
As pointed out by \citet{comeron08}, while a distance of 150 pc seems adequate for most of the clouds of the complex, a value of 200 pc 
is likely to be more appropriate for Lupus$\,$3. We therefore adopt a distance of 200 pc throughout the analysis presented in this paper.
Adopting the distance of 200 pc, and evolutionary tracks of \citet{baraffe98}, \citet{comeron03} derive an age of 1-1.5 Myr for most of their
observed members.

In this work we present new observations of the Lupus$\,$3 cloud using the MOSAIC-II optical imager and NEWFIRM near-infrared imager
at the CTIO-4m telescope. Observations and data reduction are explained in Section~\ref{Obs&DR}. Photometry, proper motions and criteria 
for candidate selection are presented in Section~\ref{candsel}, and the spectral analysis in Section~\ref{specanal}. 
The results are discussed in Section~\ref{discuss}. Finally, we summarize the main conclusions in Section~\ref{summary}.

\begin{figure*}
\centering
\resizebox{14cm}{!}{\includegraphics{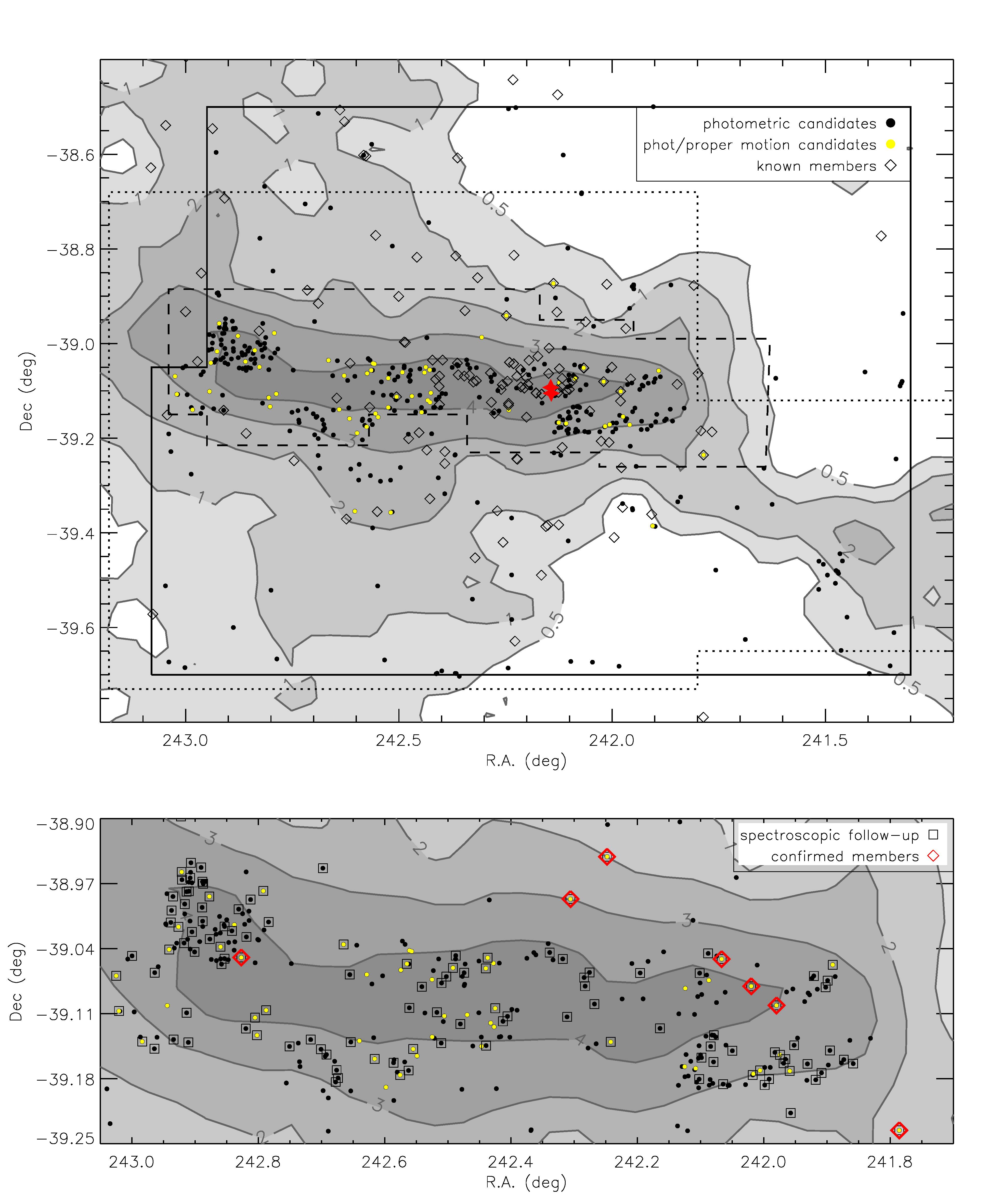}}
\caption{{\it Upper panel:} Spatial distribution of the candidate sources in Lupus$\,$3. Photometric candidates are shown as black dots, those
 also selected by proper motions are shown in yellow. Diamonds mark previously known members, with the two red stars marking
the brightest pair, HR5999/6000. Solid line shows the field-of-view of our
optical and near-infrared survey, and the dotted line the extent of the WFI field used as the first epoch for the proper motions.
The dashed lines indicate the region within which the VIMOS spectroscopic fields have been arranged. The shaded contours indicate
A$_V$ from the extinction map by \citet{cambresy99}.
{\it Lower panel:} Zoom into the region covered by spectroscopy. All the candidates observed in our follow-up are marked with squares, and
the confirmed objects are represented by red diamonds. 
}
\label{spatial}
\end{figure*}

\begin{figure*}
\centering
\resizebox{15cm}{!}{\includegraphics{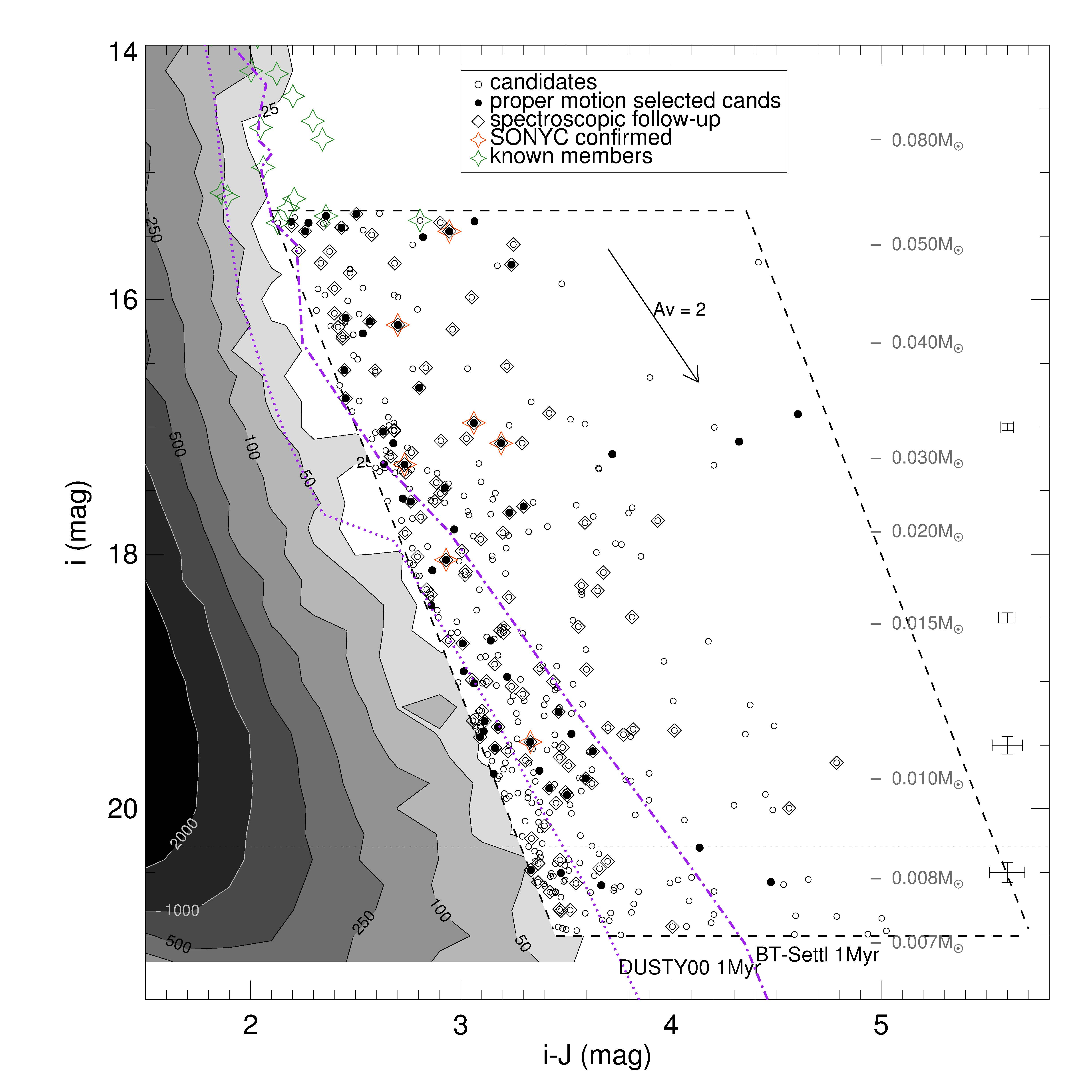}}
\caption{($i$,$i - J$) color-magnitude diagram. Open circles represent all the photometrically-selected candidates located inside
the selection box (dashed lines), 
with those additionally selected on the basis of their proper motions shown as filled circles. Diamonds 
show SONYC spectroscopy follow-up targets, with spectroscopically confirmed VLMOs highlighted as red (identified in this work) and green (previously known members) stars. 
Atmosphere model isochrones for age 1 Myr are overplotted for DUSTY00 \citep{chabrier00} and 
BT-Settl \citep{allard11}. For clarity, the sources outside the candidate selection box are represented with contours, where the 
number on each contour represents the number of sources within a 0.2 mag $\times$ 0.2 mag bin. On the right-hand side of the figure we show 1$\sigma$ uncertainties
of the photometry. The dotted line shows the completeness limit of our survey.
}
\label{IJCMD}
\end{figure*}

%--------------------------------------------------------------

\section{Observations and Data Reduction}
\label{Obs&DR}

\subsection{Optical imaging}
The imaging in the $i$-band was conducted using the MOSAIC-II CCD imager at the CTIO 4-meter telescope in Chile.
MOSAIC-II is a wide-field imager with 8 CCDs arranged in a $4\times 2$ array. It provides a field of view of 36$\farcm$8
on a side, with a pixel scale of $0\farcs27$ per pixel.
We observed four pointings arranged around the two bright Herbig Ae/Be stars. The details of the observing run are summarized 
in Table~\ref{T_obs}, and the observed field is shown in Figure~\ref{spatial}.

We reduced the MOSAIC-II images using tools in the $mscred$ package within IRAF. $Mscred$ contains the standard reduction tasks for CCD images, but adapted for the use on multi-chip mosaics. Series of dark frames and dome flats were combined to build an average zero and flatfield image. The zero and flatfield correction of the science mosaics was then carried out using the task 'ccdproc'. This also includes the subtraction of the bias level, derived from unexposed areas of the chip.
For each of the four fields, the total integration time was split in a series of 24 exposures. To coadd these exposures, we first derived matching coordinate systems for all of them, using a list of USNO-A2 stars (obtained from Vizier) as reference. The frames were then matched and stacked using the tasks 'mscimage' and 'mscstack'. Apart from a small region close to the center of the mosaic, which was excluded, this matching algorithm produces very good results. Note that the depth of the mosaics is not constant due to the small gaps between the detectors.

Photometry on the stacked images was carried out using 'daophot' tools within IRAF. After generating a source catalogue with 'daofind', the magnitudes were measured within a fixed aperture radius of 10 pixels using 'phot'. The sky background was estimated from an annulus with inner/outer radius of 10/15 pixels.
To calibrate the $i$-band photometry we compared our instrumental magnitudes with the magnitudes of sources with 
$14.5 < i < 17.0$ in Lupus$\,$3 from the {\it Deep Near Infrared Survey} catalog (DENIS; \citealt{DENIS}),
which used a Gunn $i$ filter centered at 0.8$\,\mu$m. For each field 
we calculate a zero point and apply it to all sources to create a final $i$-band catalog. 
%The data points were weighted by the uncertainties of the DENIS data. 
The color-dependence of the calibration was investigated
by calculating a coefficient $a$ in the color-term a$\times(i-J)$\footnote{$J$-band photometry from NEWFIRM data (Section~\ref{NIR_im}).}, which typically gives a$\sim$0.002. Its effect on photometry
would be much smaller than the involved photometric uncertainties, and therefore it was not taken into account.
Photometric uncertainties combine the errors returned by 'daophot', with the errors of the photometric calibration. Typical photometric uncertainties
are 0.05 mag at $i=19$ and 0.1 at $i=21$. The completeness limit of the deep exposures is $i\sim$20.3, with the magnitude range of $\sim$15.3 - 22.

For the proper motion calculation (see Section~\ref{pm}) we also used the $I_C$-band data obtained with
the Wide Field Imager (WFI) at the MPI-ESO 2.2-m telescope at La Silla Observatory \citep{baade99}, that were
previously published in \citet{lopez-marti05}. We therefore refer the reader to that paper for a detailed
description of the data reduction procedures. The field observed with WFI is outlined with dotted line in Figure~\ref{spatial}.

\begin{deluxetable}{cllllll}
\tabletypesize{\scriptsize}
\tablecaption{Observing log for our imaging and spectroscopic surveys.}
%\tablewidth{0pt}
\tablecolumns{7}
\tablehead{\colhead{field \#} & 
	   \colhead{$\alpha$(J2000)} & 
	   \colhead{$\delta$(J2000)} & 
	   \colhead{filter/grism} & 
	   \colhead{UT date} & 
	   \colhead{on-source time} & 
	   \colhead{instrument} 
}
\startdata
\multicolumn{7}{c}{\bf Imaging}\\[2ex]
\hline
1 & 16:10:10 & -38:49:08 & $i$ & 12 06 2010 & 8100$\,$s & MOSAIC-II\\
2 & 16:06:53 & -38:49:37 & $i$ & 12 06 2010 & 6900$\,$s & MOSAIC-II\\
3 & 16:10:33 & -39:22:34 & $i$ & 13 06 2010 & 7500$\,$s & MOSAIC-II\\
4 & 16:06:56 & -39:22:16 & $i$ & 13 06 2010 & 9000$\,$s & MOSAIC-II\\
\\
%1 & 16:11:01 & -38:38:20 & $J$, $K_S$ 		   & 15-16 06 2010 		  & 450$\,$s, 400$\,$s & NEWFIRM\\
%2 & 16:08:39 & -38:38:20 & $J$, $K_S$, $H1$, $H2$  & 15-16 06 2010; 05-06 05 2011 & 450$\,$s, 400$\,$s, 1900$\,$s, 1900$\,$s & NEWFIRM\\
%3 & 16:06:18 & -38:38:20 & $J$, $K_S$, $H1$, $H2$  & 15-16 06 2010; 05-06 05 2011 & 450$\,$s, 420$\,$s, 1900$\,$s, 1900$\,$s & NEWFIRM\\
%4 & 16:11:01 & -39:05:56 & $J$, $K_S$, $H1$, $H2$  & 15-16 06 2010; 05-06 05 2011 & 450$\,$s, 400$\,$s, 1900$\,$s, 1900$\,$s & NEWFIRM\\
%5 & 16:08:39 & -39:05:56 & $J$, $K_S$, $H1$, $H2$  & 15-16 06 2010; 05-06 05 2011 & 480$\,$s, 460$\,$s, 1900$\,$s, 1900$\,$s & NEWFIRM\\
%6 & 16:06:18 & -39:05:56 & $J$, $K_S$, $H1$, $H2$  & 15-16 06 2010; 05-06 05 2011 & 450$\,$s, 420$\,$s, 1900$\,$s, 1900$\,$s & NEWFIRM\\
%7 & 16:11:01 & -39:33:32 & $J$, $K_S$ 		   & 15-16 06 2010; 05-06 05 2011 & 450$\,$s, 440$\,$s & NEWFIRM\\
%8 & 16:08:39 & -39:33:32 & $J$, $K_S$, $H1$, $H2$  & 15-16 06 2010; 05-06 05 2011 & 450$\,$s, 440$\,$s, 1900$\,$s, 1900$\,$s & NEWFIRM\\
%9 & 16:06:39 & -39:33:32 & $J$, $K_S$, $H1$, $H2$  & 15-16 06 2010; 05-06 05 2011 & 450$\,$s, 460$\,$s, 1900$\,$s, 1900$\,$s & NEWFIRM\\
1 & 16:11:01 & -38:38:20 & $J$, $K_S$ & 15 06 2010, 16 06 2010 & 450$\,$s, 400$\,$s & NEWFIRM\\
2 & 16:08:39 & -38:38:20 & $J$, $K_S$ & 15 06 2010, 16 06 2010 & 450$\,$s, 400$\,$s & NEWFIRM\\
3 & 16:06:18 & -38:38:20 & $J$, $K_S$ & 15 06 2010, 05 05 2011 & 450$\,$s, 420$\,$s & NEWFIRM\\
4 & 16:11:01 & -39:05:56 & $J$, $K_S$ & 15 06 2010, 05 05 2011 & 450$\,$s, 400$\,$s & NEWFIRM\\
5 & 16:08:39 & -39:05:56 & $J$, $K_S$ & 15 06 2010, 05 05 2011 & 480$\,$s, 460$\,$s & NEWFIRM\\
6 & 16:06:18 & -39:05:56 & $J$, $K_S$ & 15 06 2010, 05 05 2011 & 450$\,$s, 420$\,$s & NEWFIRM\\
7 & 16:11:01 & -39:33:32 & $J$, $K_S$ & 15 06 2010, 05 05 2011 & 450$\,$s, 440$\,$s & NEWFIRM\\
8 & 16:08:39 & -39:33:32 & $J$, $K_S$ & 15 06 2010, 05 05 2011 & 450$\,$s, 440$\,$s & NEWFIRM\\
9 & 16:06:39 & -39:33:32 & $J$, $K_S$ & 15 06 2010, 05 05 2011 & 450$\,$s, 460$\,$s & NEWFIRM\\
\\
\multicolumn{7}{c}{\bf Spectroscopy}\\[2ex]
\hline
1 & 16:11:24 & -39:01:12 & LR\_red & 26 06 2011 & 3500$\,$s & VIMOS \\
2 & 16:11:18 & -39:01:12 & LR\_red & 12 06 2012 & 3500$\,$s & VIMOS \\
3 & 16:11:00 & -39:04:48 & LR\_red & 16 06 2012 & 3500$\,$s & VIMOS \\
4 & 16:09:30 & -39:01:12 & LR\_red & 27 06 2012 & 3500$\,$s & VIMOS \\
5 & 16:08:33 & -39:05:24 & LR\_red & 15 07 2012 & 3500$\,$s & VIMOS \\
6 & 16:07:19 & -39:07:30 & LR\_red & 17 06 2012 & 3500$\,$s & VIMOS 
\enddata
\label{T_obs}
\end{deluxetable}

%Completeness limits:
%bin 0.2  0.3   0.4   0.5 	
%I  20.35 20.30 20.05 20.40 -> 20.3
%J  18.40 18.25 18.30 18.25 -> 18.3
%K  17.60 17.50 17.70 17.50 -> 17.6

\subsection{Near-infrared imaging}
\label{NIR_im}
The near-infrared observations were designed to provide $J$- and $K_S$-band photometry in the area slightly larger than the
one covered with MOSAIC-II. 
%The same area was covered in intermediate-band filters H1 and H2 \citep{vandokkum09}, centered at
%1.560$\,\mu$m and 1.708$\,\mu$m, respectively, and  
%each covering approximately half of the standard H-band broad-band filter.  
We used NEWFIRM at the CTIO 4-m telescope, providing a field of view of $28'\times 28'$ and a pixel scale of 0\farcs4. 
The NEWFIRM detector is a mosaic of four $2048\times2048$ Orion InSb arrays, organized in a $2\times2$ grid.  
See Table~\ref{T_obs} for the details of the NEWFIRM observing runs. 
Data reduction was performed using the NEWFIRM pipeline. The data were dark- and sky-subtracted, corrected for 
bad pixels, and flat field effects. The astrometry was calibrated with respect to the {\it Two 
Micron All Sky Survey} (2MASS; \citealt{2mass}) coordinate
system, and each detector quadrant was re-projected onto an undistorted celestial tangent plane. 
The four quadrants were then combined into a single image, followed by the final
stacking into mosaics.
Source extraction was performed using {\em SExtractor}, requiring 
at least 5 pixels with flux above the $3\sigma$ detection limit,
followed by the rejection of sources at the edges of the detector
and overly elongated objects ($a/b\,>2.0$).
Photometric zero-points for each NEWFIRM field were calculated by matching the sources in our catalog with the sources found in 2MASS. 
The completeness limits of the J- and K$_S$-band catalogues are at J=18.3 mag and K$_S$=17.6. Typical photometric uncertainties are 
$\leq 0.05$ mag at J$\leq18$ and K$_S\leq 17$, 
and 0.1 mag at J=19 and K$_S$=18.

%Photometric calibration was done by matching the sources in our catalog with the sources found in 2MASS. 
%NEWFIRM broad-band filters are designed in the MKO system \citep{tokunaga02}. We applied the transformation
%between the MKO and 2MASS phtometric system as given in the Explanatory Supplement to the 2MASS All Sky Data Release 
%\footnote{http://www.ipac.caltech.edu/2mass/releases/allsky/doc/sec6\_4b.html}. 

\subsection{Multi-object spectroscopy}
We obtained optical spectra using VIMOS/VLT in the Multi-Object Spectroscopy (MOS) mode in programs 087.C-0386 and 089.C-0432.
One field was observed in P87 and 5 in P89\footnote{ESO's period P87 denotes a period Apr - Sep 2011, and P89 Apr - Sep 2012}.
VIMOS is a wide-field imager with 4 CCDs arranged in a $2\times 2$ array. Each detector covers  a field of 
view of $7'\times8'$ with a pixel resolution of 0\farcs205. The four quadrants are separated by gaps of 
approximately $2'$. The spectra were obtained  
using the low resolution red grism (LR$\_$red). We covered the wavelength regime between 5500~\AA~and 9500~\AA, 
with the dispersion of 7.1\AA~per pixel and slit width of 1$''$, 
resulting in a spectral resolution of R$\,\sim$210. The total on-source time for each field was $\sim 3500\,$s. 
The dashed line in Figure~\ref{spatial} outlines the area where the VIMOS fields have been arranged.

Data reduction was performed using the VIMOS pipeline provided by ESO. Data reduction steps include bias subtraction, 
flat-field and bad-pixel correction, wavelength and flux calibration, as well as the final extraction of the spectra.
The extraction is done by applying the optimal extraction from \citet{horne86}, which averages the signal optimally weighted by a function of the 
signal noise. 
In total we obtained 124 spectra. For the description of the selection criteria for spectroscopy, see Section~\ref{candsel}.
%We obtained 124 spectra of the objects inside the candidate selection box, including 26 that were also
%proper-motion selected. 

%-------------------------------------------
\section{Candidate selection}
\label{candsel}

We start this section with a summary of the candidate selection procedure. All individual steps are in detail
explained in the following subsections.

\begin{itemize}
\item  Our parent sample comprises the $\sim500,000$ sources in the $iJ$ photometric
catalog created from the MOSAIC and NEWFIRM observations.
\item $(i, i-J)$ photometric selection results in 409 candidates (see Section~\ref{CMD_sel}).
\item  We further narrowed the sample of $(i, i-J)$ photometric candidates from 409 to 59, by applying the proper motion selection criterion (see Section~\ref{pm}).
In the following sections, we refer to this sample as ``$IJ$-pm'' sample.
\item To complement our main optical-NIR selection, we selected mid-infrared excess sources using $Spitzer$ IRAC colors (75 candidates). This list contains 19 proper motion candidates (see Section~\ref{MIR}). We refer to this sample as ``IRAC-pm'' sample.
\end{itemize}

Section~\ref{specanal} describes the
selection of targets for spectroscopic follow-up from the above mentioned
candidate lists.

\subsection{Optical and near-infrared photometric selection}
\label{CMD_sel}

Figure~\ref{IJCMD} shows the ($i$, $i-J$) color-magnitude diagram of the sources observed in Lupus 3.
%Previously spectroscopically confirmed M-type members are shown as XXX. 
%On the right-hand side of the diagram, we plot 
%the expected $I$-band brightness of Lupus$\,$3 members with the mass 0.075, 0.015, and 0.005 $M_{\odot}$, each at $A_V$=0 and 5. 
%For the calculation we use  COND03 \citep{baraffe03}, DUSTY00 \citep{chabrier00} and BT-Settl \citep{allard11} evolutionary models. 
Very-low-mass (VLM) objects are expected to occupy a distinct region on the red side of the broad cumulation of the background 
main-sequence stars. As in all the previous SONYC works, we construct the selection box in the CMD such that it represents a continuation of the sequence of the known members. 
We note that the majority of the known members is brighter than $i$=15.5, and are thus located outside the range of the CMD shown in Figure~\ref{IJCMD} (the saturation level of our optical dataset is around $i=15.3\,$mag). 
Green stars show a subset of known spectroscopic members from \citet{mortier11,comeron13} fainter than $i$=14.0, with $i$ and $J$ photometry from DENIS and 2MASS,
respectively. 
%As the transformation between Cousins I (most of the known members have $I_C$ available in the literature) and the Gunn-$i$ filter is not
%well established, we prefer not to show the previously known members in the same plot. For this reason
For clarity, below $i=15.3\,$mag we show only the three members that overlap with the sources in our selection box (green stars).

For comparison we show DUSTY00 \citep{chabrier00} and BT-Settl \citep{allard11} 
isochrones for 1$\,$Myr, adjusted to the distance of Lupus$\,$3.
The isochrones\footnote{available at \url{http://phoenix.ens-lyon.fr/Grids/}} are shown in the same photometric filters (DENIS and 2MASS) as our data.
According to the BT-Settl isochrone, the majority of the low mass members are expected to be within our selection box, while looking at DUSTY we might be 
missing some objects with masses above 0.02$\,$M$_{\odot}$. However, it has to be taken into account that the isochrones have no extinction applied to them, while the potential members of Lupus are expected to have a certain degree of extinction, i.e. we expect them to be located to the right of the 
isochrones. 
%The observed field towards Lupus$\,$3 is very crowded, and one can expect high degrees 
%of contamination by background sources. Shifting the border of the selection box to the left quickly doubles the candidate list and 
%makes the spectroscopic follow-up extremely time-consuming. The final position of the selection box is therefore a trade-off between an unbiased 
%photometric sample, and an efficient spectroscopic follow-up. 

%While the BT-Settl grid is readily available in the DENIS and the 2MASS filters, DUSTY and COND
%have to be converted into the appropriate photometric system. No transformation was applied to the Cousins I-band
%magnitudes since the DENIS system appers to be close enough ($\leq 0.05$mag; \citealt{delfosse97, costa06}), and the 
%J-band was converted to 2MASS system using transformations
%given in \citet{carpenter01}. 

In total, 409 sources were found within the selection box and served as the initial list of 
candidates for spectroscopic follow-up with VIMOS. Further selection was performed based on the proper-motion measurements, as
described in the following section.

\begin{figure*}
\centering
\resizebox{14cm}{!}{\includegraphics{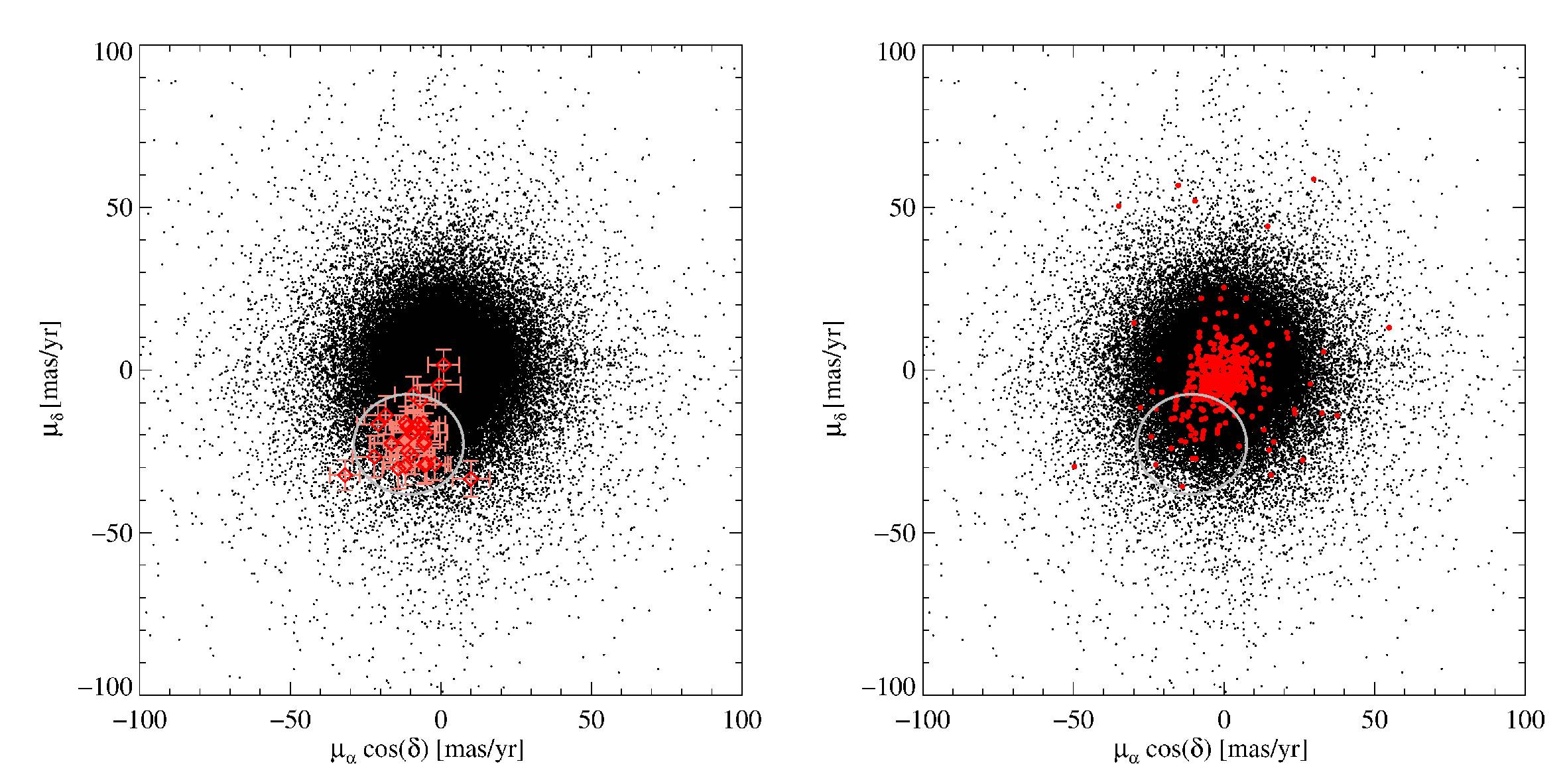}}
\caption{Proper motions of objects in the Lupus$\,$3 field, based on 11-12$\,$yr baseline between the WFI and NEWFIRM data (black dots). 
{\it Left:} The red symbols show the measured proper motions for the sources identified as members in \citet{lopez-marti11}. The ellipse depicts the $2\sigma$ 
criterion used for the proper motion candidate selection (see Section~\ref{pm}). 
{\it Right:} Red dots mark the proper motions of our 409 candidates selected from optical and NIR photometry. 
There are 59 candidates that pass both photometric and proper motion
selection criteria.}
\label{pm_all}
\end{figure*}

\subsection{Proper motions}
\label{pm}

%\begin{figure}
%\centering
%\resizebox{9cm}{!}{\includegraphics{pm_comparison_2_4.jpg}}
%\caption{Comparison of the proper motions of the Lupus$\,$3 members from \citet{lopez-marti11} (LM11; black circles) and the proper
%motions of the same objects calculated in this work (red diamonds). Both ellipses are centered at the average proper motion of the Lupus$\,$3 members
%\citep{lopez-marti11}, with the semi-major axes equivalent to $2\sigma$ (dashed line) and $3\sigma$ (full line) uncertainties. 
%The $2\sigma$ ellipse was used as a criterion for the candidate selection described in text.}
%\label{pm_comp}
%\end{figure}

\begin{figure*}
\centering
\resizebox{14cm}{!}{\includegraphics{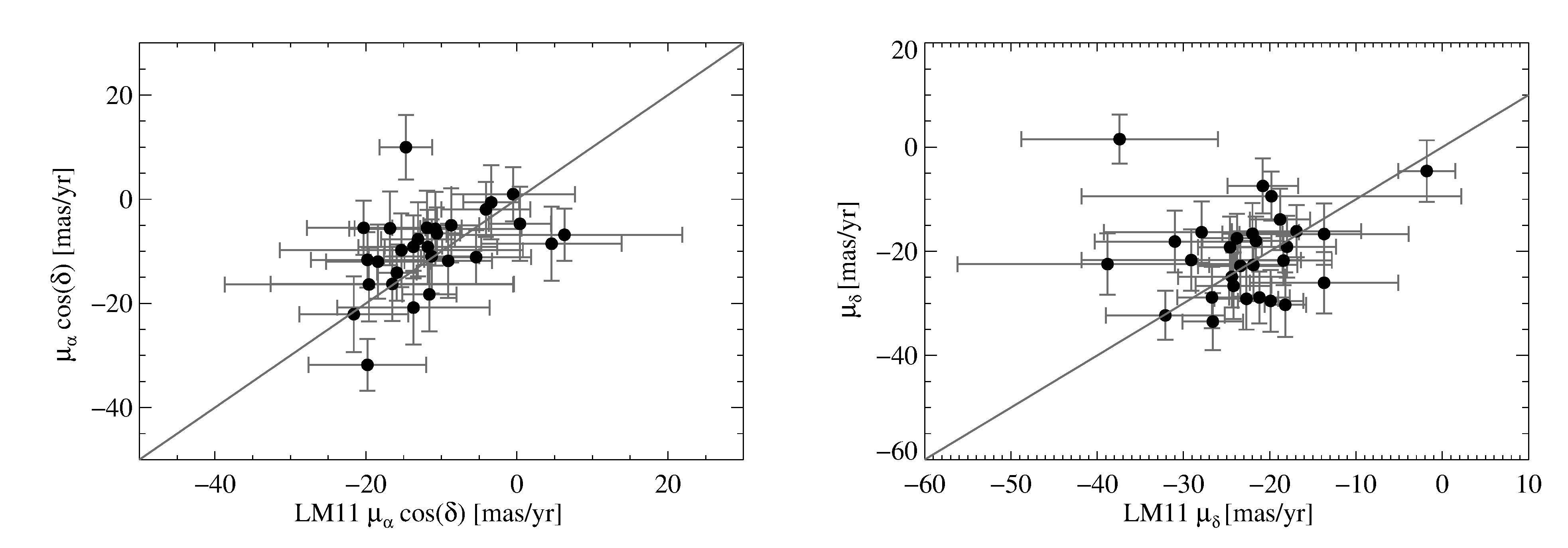}}
\caption{Comparison of the proper motions of the Lupus$\,$3 members from \citet{lopez-marti11} and this work. 
A line $x=y$ is shown as a solid line.} 
%The $3\sigma$ outlier is marked with red.}
\label{pm_comp_radec}
\end{figure*}

Proper motion measurements are based on the WFI dataset obtained between May 28th and June 3rd 1999 and our
NEWFIRM data from June 2010 and May 2011. Whenever possible, we 
use the data from 2011, to assure the longest time baseline.
In the following we describe in detail all the steps performed:\\
(1) The overlapping fields were first divided into sub-regions with the maximum size of $3' \times 3'$, and 
with an overlap of 0\farcm5 between
the adjacent fields. We find that an area of this size typically contains a high number of sources that can be used 
for astrometric transformations, while being reasonably small to minimize the effects of field distortions.\\
(2) The absolute coordinate calibration shows systematic offsets of up to $12''$ between epochs in some regions, which poses
 a problem with cross-referencing individual sources, given the on-sky source density in Lupus$\,$3. It is therefore crucial 
to apply an initial offset to the coordinates in one of the epochs, to assure the correct
source registration in the two epochs. To determine the offset for each sub-region created in the first step, we 
create two images of the same size as the NEWFIRM and WFI sub-region of interest. The images have value of 1 
at the positions equivalent to the centroid pixels of bright stars, and zero elsewhere. % in overlapping regions, and 
These images are then cross-correlated to obtain an average shift applied to a particular region;\\
(3) In this step we cross-match the catalogs based on right ascension and declination. This step is used only to identify individual sources in the two epochs, 
but not to calculate proper motions. The maximum matching radius of $2''$ was used, which is equivalent $\sim170\,$mas\,yr$^{-1}$ over the 11-12 yr baseline.\\
(4) We apply a cutoff in magnitude for the matching sources, in order to discard saturated and faint ones (we keep all
the sources with $J\approx15.5 - 17.5$). The remaining sources are used in the next step to
calculate a transformation of WFI positions to NEWFIRM coordinates. 
The number of sources used for the transformation in each sub-region
varies between 80 and 400, with an average of 240; \\
(5) In the final step we calculate a polynomial that describes the transformation of the WFI detector ($x,y$) positions into the NEWFIRM
coordinate frame for each sub-region, using the IDL procedure POLYWARP.
In order to avoid the uncertainties in the absolute 
coordinate calibration, this step was performed in detector coordinates relative from one dataset to the other. 
We tested a 2nd and a 3rd degree polynomial and find that the two transformations give very similar final results, 
which is expected given the large number of sources available to calculate the transformation.
Any polynomial with a degree $\geq 2$ corrects for shift and rotation between the fields, and also largely compensates for the 
distortions of the two detectors (e.g. \citealt{schoedel09}). After applying the transformation, a proper motion was calculated for each source.
Since the vast majority of stars used to calculate the transformation between the epochs are not Lupus members, we expect a
circularly symmetric distribution around (0,0) in the proper motion space. We note that some of the members might have been included in the 
transformation, but their number compared to the total number of sources in our survey area is negligible, and thus does not influence
the results.
Finally, proper motions for the sources towards Lupus$\,$3 are shown as black dots in Figure~\ref{pm_all}.

%The procedure POLYWARP that was used to calculate the coefficients of the transformation, unfortunately does not supply the 
%uncertainties of the polynomial fit. 
In order to assess
the uncertainties of the proper motion measurements, we use the fact that the chips were divided into overlapping sub-regions in step (1) described above. This allows
us to derive the uncertainties by comparing proper motions of common sources that were obtained using different
sets of reference stars. 
For each NEWFIRM field, we adopt a common uncertainty calculated as a standard deviation of all the differences 
obtained from the overlapping regions in each field. For the sources with more than one measurement, the average
is adopted as the final proper motion value.

\citet{lopez-marti11} performed a kinematic study of the bright objects in the  Lupus regions and found the average
%compiled a list of members of Lupus 1, 3 and 4, and calculated their proper motions from
%the positions in various publicly available catalogs.
%Virtual
%Observatory tools\footnote{\url{http://www.ivoa.net}}. 
proper motion of the Lupus$\,$3 members to be $v_{\alpha}=-10.7\pm9.1\,$mas$\,$yr$^{-1}$, 
and $v_{\delta}=-22.8\pm7.7\,$mas$\,$yr$^{-1}$. The uncertainties quoted here represent the standard deviation of the measurements
for Lupus$\,$3 given in Table 2 of \citet{lopez-marti11}.
We select all the sources within an ellipse centered at (-10.7, -22.8)$\,$mas$\,$yr$^{-1}$ and with the semi-major axes equal to two times
the above quoted uncertainties. This list was cross-matched with the photometric candidate catalogue described in the previous section, 
resulting finally in the list of 59 high-priority candidate members (``$IJ$-pm'' sample; shown as filled black circles in Figure~\ref{IJCMD}).
The proper motions of the members in common with \citet{lopez-marti11} are shown as red symbols in the left panel of Figure~\ref{pm_all},
on top of the circular proper motion distribution of all other sources in our field (black dots). The right panel shows
the proper motions of the photometric candidates identified in this survey (red points).

%Figure~\ref{pm_comp} shows a comparison between the proper motions of the Lupus$\,$3 members from \citet[black circles]{lopez-marti11}
%and the proper motions of these same sources obtained in this work (shown as red diamonds). The agreement between the two results is very good --
%90\% of the proper motions agree within $3\sigma$.
Figure~\ref{pm_comp_radec} shows a comparison between the proper motions of the Lupus$\,$3 members from \citet{lopez-marti11} 
and the proper motions of these same sources obtained in this work. There are in total 40 matching objects. 
While virtually all of these sources are saturated in WFI images, the brightest ones show severe saturation effects such as strong CCD leaking, 
which can seriously affect the astrometry. We therefore discard the extremely saturated sources with I$_C > 12.2$.
%Open red circles mark the six objects whose proper motions disagree by
%more than $3\sigma$. 
%A closer inspection of the outliers shows that five of them are among the six brightest sources in the sample of 40 examined objects. 
%While virtually all of these sources are saturated in WFI images, only the brightest ones show severe saturation effects such as strong CCD leaking, 
%which most probably seriously affect our astrometry. This might be a reason why the disagreement between our proper motions and those from \citet{lopez-marti11}
%is stronger in declination, which matches the direction of the leak columns in the WFI CCD.
%Among the remaining 30, 
%there is only one object whose proper motion disagrees by more than $3\sigma$. 
%It is not clear where the discrepancy comes from, since the object is $\sim$3 magnitudes fainter than the excluded problematic sources, 
%i.e. the saturation should not have a strong effect on the astrometry. Visual inspection of this object in the NEWFIRM and WFI images does not reveal any
%obvious defect. 
The agreement between the two sets of proper motions is satisfactory, with all proper motions agreeing within $3\sigma$.
%$85\%$ of all the proper motions agreeing within $3\sigma$. Excluding the 8-10 brightest sources (I$<12.1-12.2$) from the comparison sample,
%more than $95\%$ (29/30) proper motions agreeing within uncertainties.

\subsection{Mid-infrared selection}
\label{MIR}
% spitzer 75 candidates, 41 with I>15.3 and in IJ catalog
% 7 in the selection box, 3 observed, all 3 confirmed
% 11 allocated from Spitzer, 3 in box
% remaining 8 are NOT on the PM selected list.

We retrieved the Spitzer IRAC and MIPS photometry of Lupus$\,$3 from the Legacy Program data archive available at the
Spitzer Science Center, using the ``High reliability catalog'' (HREL) created by the ``Cores to Disk'' (c2d) Legacy team,
made available at the SSC Web site \footnote{{\tt http://data.spitzer.caltech.edu/popular/c2d/20071101$\_$enhanced$\_$v1/lupus/catalogs/}}.

\begin{figure}
\centering
\resizebox{9cm}{!}{\includegraphics{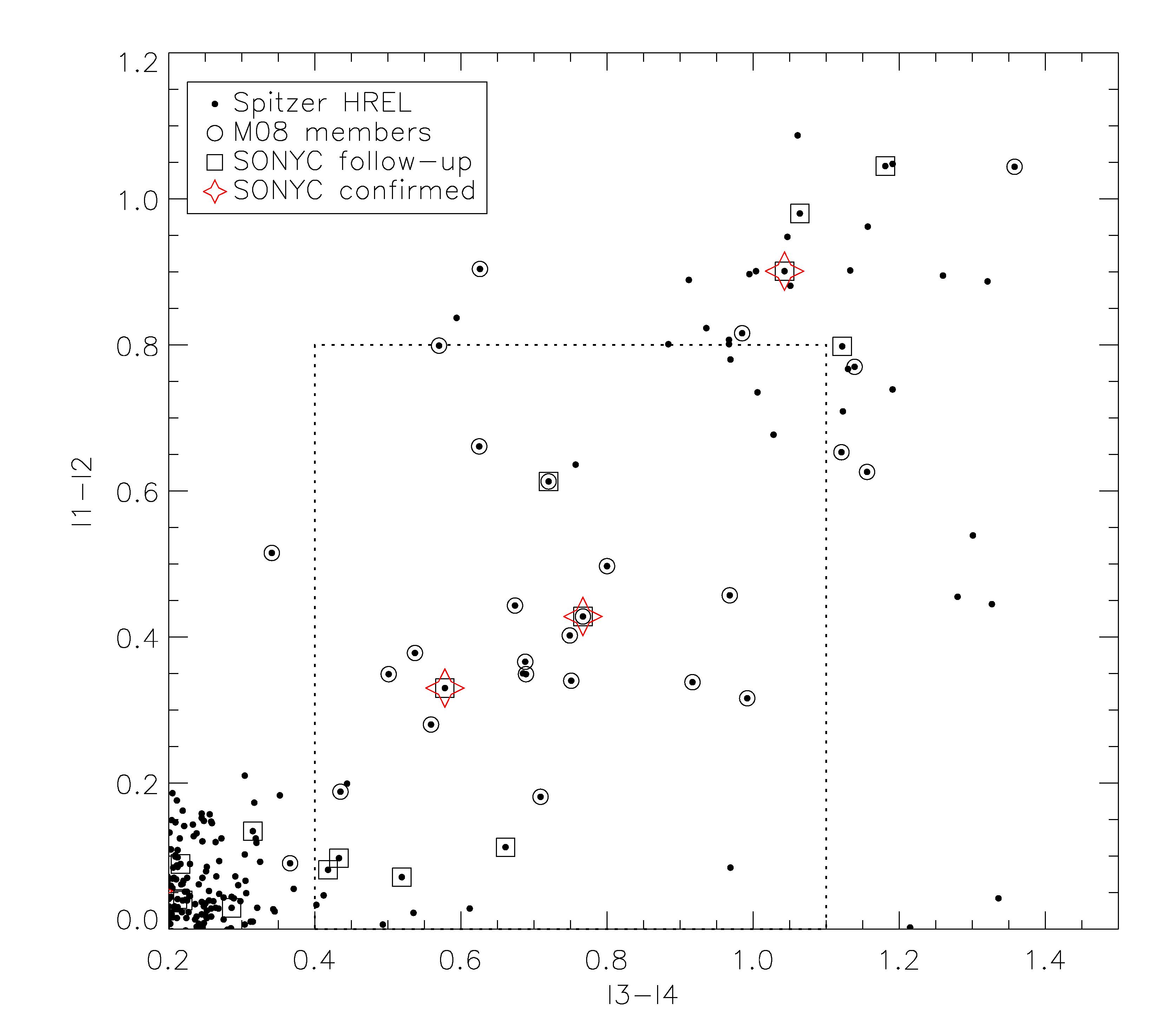}}
\caption{[3.6]-[4.5], [5.8]-[8.0] diagram constructed from $Spitzer$ IRAC photometry, for the objects matched with
our optical catalog. The dashed line denotes the area where Class II objects are located, based on \citet{allen04}. 
Open circles mark the collection of YSOs and pre-main sequence stars from \citet{merin08}. Squares mark the objects whose
spectra were obtained as part of the SONYC follow-up, and the red stars mark the three confirmed VLMOs.
}
\label{MIR_Fig}
\end{figure}

The $Spitzer$ catalogue was cross-correlated with our $i$-band catalogue, requiring the separation of $2''$ or better, and
uncertainties $\leq 0.2\,$mag in the first four IRAC bands. To select the MIR-excess sources, we require $I3-I4 \geq 0.4$ and
$I1-I2 > 0$. This selection contains 75 sources, shown in Figure~\ref{MIR_Fig} as dots to the right of $I3-I4 \geq 0.4$ line.
Squares mark the 11 sources selected for follow-up spectroscopy.
Of the 75 Spitzer-IRAC candidates, there are 19 sources that satisfy the proper motion criterion applied to the $iJ$ candidate list, and were included
in the high-priority candidate list supplied to the VIMOS allocation software (we name this list the ``IRAC-pm'' sample). 
%The remaining sources are either outside of the overlap
%area between SONYC and WFI survey, not detected in optical and/or NIR images, or afected by artefacts.
The Spitzer-IRAC list and the $iJ$ photometric candidate list have 7 objects in common.

%-----------------------------------------------------------------------------------------

\section{Spectral analysis}
\label{specanal}
%\begin{figure*}
%\centering
%\resizebox{18cm}{!}{\includegraphics{Na_grav_spectra.jpg}}
%\caption{Region around the Na~I gravity-sensitive absorption feature at $\sim8200\,$\AA. The average wavelength of the Na~I doublet
%is marked by the vertical line. We show a spectral sequence M3 to M7, and a comparison of young cluster member spectra with those of giants and
%field dwarfs. All the spectra have been smoothed to match the spectral resolution of VIMOS, and normalized at 8260$\,$\AA.
%}
%\label{gravity_spec}
%\end{figure*}

%\begin{figure*}
%\centering
%\resizebox{18cm}{!}{\includegraphics{Na_grav_models.jpg}}
%\caption{Region around the Na~I gravity-sensitive absorption feature at $\sim8200\,$\AA. The average wavelength of the Na~I doublet
%is marked by the vertical line. We show a model spectral sequence between 3400 K and 2900 K, approximately
%equivalent to spectral types M3 to M7 shown in Figure~\ref{gravity_spec}, at three values of log$\,g\,$. BT-Settl \citep{allard11} have 
%been smoothed to match the spectral resolution of VIMOS, and normalized at 8260$\,$\AA.
%}
%\label{gravity_models}
%\end{figure*}

%\begin{figure*}
%\centering
%resizebox{18cm}{!}{\includegraphics{Na_grav_examples.jpg}}
%\caption{Region around the Na~I gravity-sensitive absorption feature at $\sim8200\,$\AA. The spectra of three objects with different
%effective temperatures (2600$\,$K, 2900$\,$K, and 3500$\,$K; black) are overplotted by the model spectra at different gravities (pink).}
%\end{figure*}

\begin{figure}
\centering
\resizebox{9cm}{!}{\includegraphics{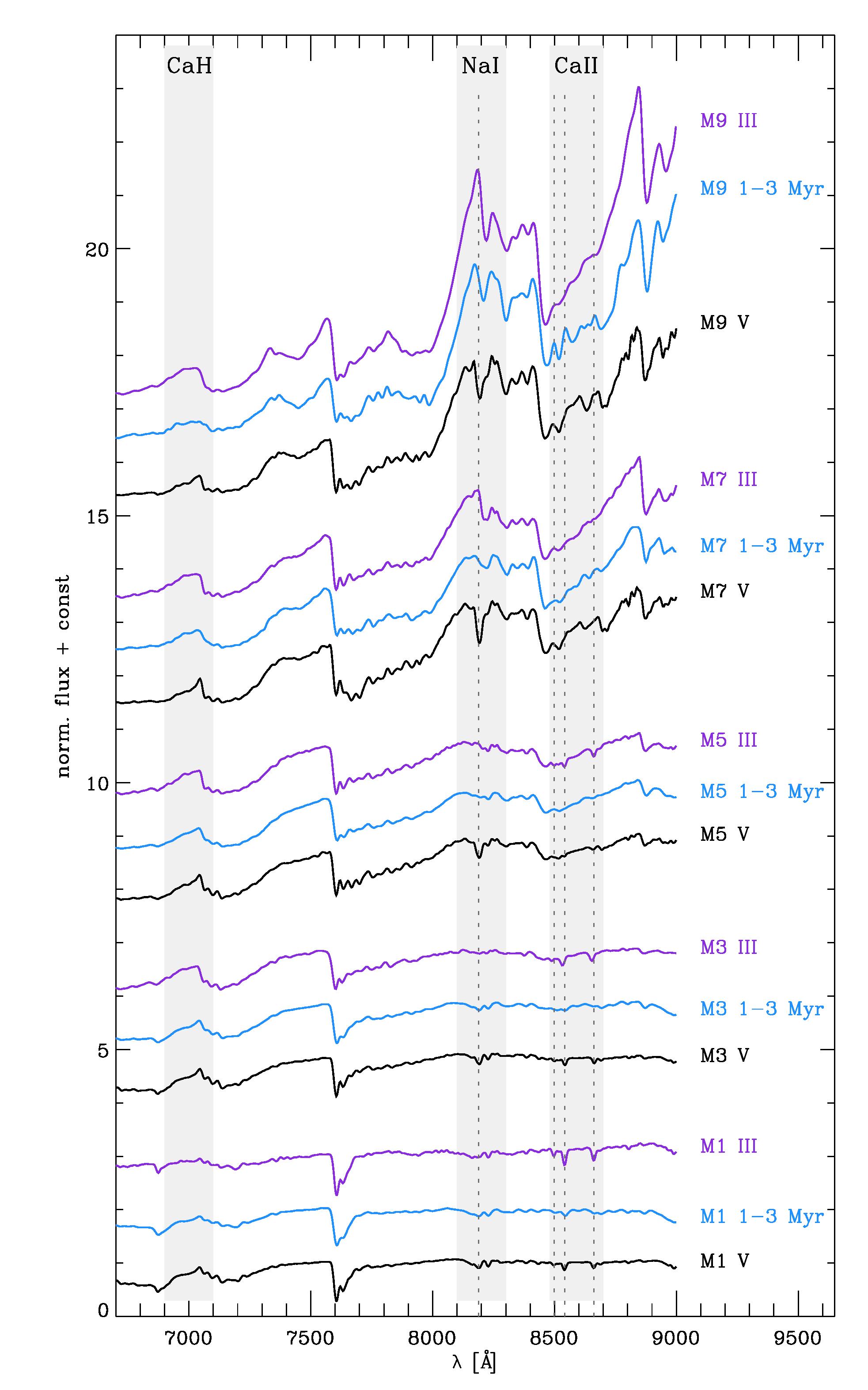}}
\caption{Part of the spectral sequence used for the spectral type fitting. Templates include
dwarf and giant stars with spectral type M1 to M9, as well as the young members of star forming regions.  
Three gravity-sensitive spectral regions are shaded: (1) the 7000\AA~region affected by CaH absorption,
(2) the region around Na~I absorption feature at $\sim8200\,$\AA, and (3) the CaII absorption triplet at $\sim8600\,$\AA.
The average wavelength of the Na~I doublet, and the CaII triplet are marked by the vertical dashed lines.
The spectra have been smoothed to match the spectral resolution of VIMOS, and normalized at 7500$\,$\AA.
}
\label{Spt_seq}
\end{figure}

% *** before 1st revision
%124 spectra in total (P89 and P87)
%26/123 proper motion selected
%In P87 allocated match with the IJ catalog I>15.3 -> 37 objects (2 in the box)

% spitzer 75 candidates, 41 with I>15.3 and in IJ catalog
% 7 in the selection box, 3 observed, all 3 confirmed
% 11 allocated from Spitzer, 3 in box
% remaining 8 are NOT on the PM selected list.

%outside the box: 35 from P87 and 8 from spitzer = 43, and +1 random (217_2)
%35 from P87 -> 2 are in the pm list

%Here: total number of allocated objects. How many from optical, how many from Spitzer?

%*** after 1st revision (July 2013)
% 409 candidates, 59/409 from pm
% 123 spectra in total (P89 and P87) + 3 from other people (green stars in CMD) = 126
% 33 spec from pm (32 our + 1 other works)
% confirmed: total 10 (7 SONYC + 3 other), 8/10 from pm (7 SONYC +1 other), 2/10 not in pm list (0 SONYC + 2 other)
Slits were allocated to 123 out of 409 objects, selected from the optical CMD,
using the VIMOS Mask Preparation Software (VMMPS). 
VMMPS allows for only two priority classes. In P89 the high priority list included the two
proper-motion selected samples, thus containing 59 candidates from the ``$IJ$-pm'' sample,
%proper motion selected candidates from the $iJ$ selection, 
%and the 19 proper motion selected candidates from the $i-Spitzer$ selection 
and 19 candidates from the ``IRAC-pm'' sample, while the rest were
the photometrically-selected candidates from $IJ$ and $Spitzer$-IRAC lists. 
In P87, prior to our proper motion analysis, the high priority candidates were those from the selection
box, and the rest were randomly chosen objects within the FOV.
In total, we obtained spectra of 123 objects from the selection box, and 44 outside of it.
Of the 123 spectra from the $IJ$ selection box, 32 are from the ``$IJ$-pm'' sample.
%are those of the proper-motion- and photometry-selected candidates.
Diamonds in Figure $\,$\ref{IJCMD} mark all the candidates with spectroscopy inside our selection box.

\subsection{Preliminary selection}

Our goal is to identify young, low mass stellar and substellar members of Lupus$\,$3. 
Among the photometrically selected candidates, we may find contaminants such as
embedded stellar members of Lupus$\,$3 with higher masses, reddened background M-type stars, and
(less likely) late M- and early L-type objects in the foreground.
The red portion of M-dwarf optical spectra is dominated by molecular features (mainly TiO and VO; \citealt{kirkpatrick91,kirkpatrick95}), 
which allows a relatively simple preliminary selection of candidates based on visual inspection. Here we included all the objects with clear evidence, 
but also, for the sake of caution, those objects with only tentative evidence of molecular features. 
Another property that was used in the selection is the slope, 
which should be positive in the red portion of the optical regime for M-type stars. 
Earlier stellar types typically have negative slopes, or appear flat (late K-type). Among the rejected spectra we mainly find flat spectra with 
no, or with very weak indication of molecular features (late K-dwarfs, or reddened earlier-type stars), and
reddened featureless spectra (background or embedded stars of spectral type earlier than K). We also looked for the 
H$_{\alpha}$ emission at 6563 \AA; only in one case an object was selected solely on the account of showing H$_{\alpha}$, and despite the spectrum showing no
prominent features characteristic for the M-type stars.
After this preliminary selection, our sample contains 27 spectra. 26 of the selected spectra are photometric candidates, while 1 source appears
bluer than expected for Lupus members. This source, SONYC-Lup3-6, will discussed in more detail in Section~\ref{S-L3-6}.

\subsection{Spectral features as membership indicators}

VIMOS was used in the spectroscopic part of the SONYC campaign in Cha-I \citep{muzic11}.
The analysis presented here is partly similar to the one in that paper.
Since the Cha-I observations in the first half of 2009, VIMOS was equipped with new, more red-sensitive detectors, which substantially
decreased fringing that previously contaminated spectra longwards of 8000\,\AA. This is fortunate not only because our
wavelength range is increased, but also because it allows us to analyze the gravity-sensitive Na~I features at $\sim\,$8200\,\AA, 
which can provide an important constraint for the age of each observed object.

Youth provides 
strong evidence for cluster membership. We use the following features 
to identify members of Lupus$\,$3:\\
(1) Na~I doublet at 8183/8195 \,\AA, which shows a change in equivalent widths, 
and a strong increase in the depth of absorption with increasing the surface gravity (e.g. \citealt{martin99, riddick07,schlieder12}).
%The alkali metal lines 
%are highly gravity-sensitive, and thus very useful for distinguishing pre-main-sequence stars from field dwarfs
%(e.g. \citealt{martin99, riddick07}). One of the strongest gravity-sensitive features in the optical is the 
%Na~I doublet at 8183/8195 \,\AA, which shows a change in equivalent widths, and a strong increase in the depth of absorption 
%with increasing the surface gravity. 
Although the individual lines of the doublet cannot be resolved in our spectra, 
the overall strength of the feature can still be used to assess the membership in the cluster even at low spectral resolutions.\\
(2) The spectral region $\sim 6900 - 7100$\AA, shaped by TiO and CaH bands, also strongly affected by gravity at M types \citep{kirkpatrick91}.
In particular, CaH absorption shortward of 7050$\,$\AA~is stronger in M-type dwarfs, making the peak of the feature appear sharper in dwarfs, and rather blunt in giants.\\
(3) H$_{\alpha}$ emission, which is usually interpreted as a sign of accretion, and thus can be used as additional evidence for youth. \\
(4) The CaII triplet lines in absorption at 8498\AA, 8542\AA, and 8662\AA~ are prominent in late-K through mid-M giants, and are weaker in dwarfs \citep{kirkpatrick91}. This feature
can help in identifying giant contaminants at early-M spectral types where Na~I doublet becomes of limited use.

The points (1), (2) and (4) are demonstrated in Figure~\ref{Spt_seq}, showing a part of the spectral sequence (M1 to M9 with a step of 2 spectral subtypes) of giants, dwarfs, and members of young star forming regions
(see Section~\ref{spt_fit} for detailed information on spectra and citations).
The sodium absorption is much stronger in field dwarfs when compared to low-gravity atmospheres, and thus can be 
used to discard background field dwarfs with SpT later then M4 from
our sample. Also, in the low-resolution spectra of dwarfs, the absorption feature is deepest at about the average 
wavelength of the Na~I doublet (vertical line in Figure~\ref{Spt_seq}), and appears symmetric around it. On the other hand, the feature appears
asymmetric in giants with the minimum slightly shifted towards longer wavelengths. This property can be used to distinguish between giants and
low-gravity young objects at spectral types later than $\sim$M6. 
Determination of the spectral class is less straightforward at early spectral types (M2-M3), where the CaH feature at
7000\AA~and Ca~II triplet can provide additional help. At M1 and earlier, 
the low-resolution spectra start to be of limited value for membership determination.   

The equivalent width (EW) of H$_{\alpha}$ line in emission is often used to distinguish the accreting objects from the non-accretors.
While the chromospheric H$_{\alpha}$ emission is expected to be found in most pre-main sequence M-dwarfs due to magnetic activity, larger
equivalent widths observed in Class II objects are usually interpreted as a consequence of accretion and/or outflows
(e.g. \citealt{scholz07, stelzer13}).
\citet{white&basri03} propose the EW(H$_{\alpha}$) for actively accreting stars to be larger than 3\AA\ for K0 -- K5 stars, 
$>$10\AA\ for K7 -- M2.5, $> 20$\AA\ for M3 -- M5.5, and $>40$\AA\ for M6 -- M7.5 stars. A similar criterion, but with finer division with respect to spectral type, and 
extended to the substellar
regime, was derived by \citet{barrado03}. The EW(H$_{\alpha}$) for the SONYC objects in Lupus$\,$3 are given in Table~\ref{T_spec}. 
SONYC-Lup3-1, 7, and 10 are revealed as strong accretors, and SONYC-Lup3-2, 8, and 11 are clearly non-accretors. 
The remaining objects with measured EW(H$_{\alpha}$) all appear to be undergoing accretion according to the \citet{barrado03} criterion, whereas
SONYC-Lup3-5, 6, and 12 fall slightly below the thresholds set by \citet{white&basri03} for their respective spectral types.

%However, the differences between young cluster members with log$\,g\,=3.5-4$ and giants (log$\,g\,\leq3$) are more subtle, and additional
%spectral signatures should be used to distinguish between those. This is discussed below in the text.

%\subsection{Spectral fitting}
%\label{fit_sec}

%In the previous papers of the SONYC series, the effective temperature ($T_{\mathrm{eff}}$) of the objects
%and membership (youth) were assessed by the method of spectral model fitting. 
%We found this method suitable for the low-resolution near-infrared observations where 
%the broad peak caused by water absorption at the edges of the H-band allows to unambiguously differentiate between 
%young, low-gravity objects and field dwarf contaminants.
%The model-fitting was again our first
%approach, but it turns out that the analysis of the gravity-sensitive features in the optical becomes much more 
%straightforward and reliable when comparing the spectra to the spectral templates of field dwarfs, giants, and young members of star forming regions. This is
%particularly true for the 7000\AA~  
%region, that is providing a progressively poorer match to the spectra as one moves towards lower temperatures.   
%In the following sections we present the new empirical fitting scheme based on the template spectra, as well as the best-model fits that will
%be used to additionally constrain $T_{\mathrm{eff}}$, and provide a comparison of different models to the data.

\subsection{Spectral classification}
\label{spt_fit}

%\begin{figure*}
%\centering
%\resizebox{16cm}{!}{\includegraphics{members_spectra_tmplt.jpg}}
%\caption{Spectra of the objects in Table~\ref{T_spec} (black), along with the best-fit spectra of members of young star forming regions (red dashed line; Luhman et al.).
%The spectral resolution of the fit templates has been reduced 
%in order to match the resolution of our spectra. Each spectrum has been corrected
%for extinction (see Table~\ref{T_spec}), but not for the atmospheric absorption.
%}
%\label{memb_spec_tmpl}
%\end{figure*}

\begin{figure}
\centering
\resizebox{9cm}{!}{\includegraphics{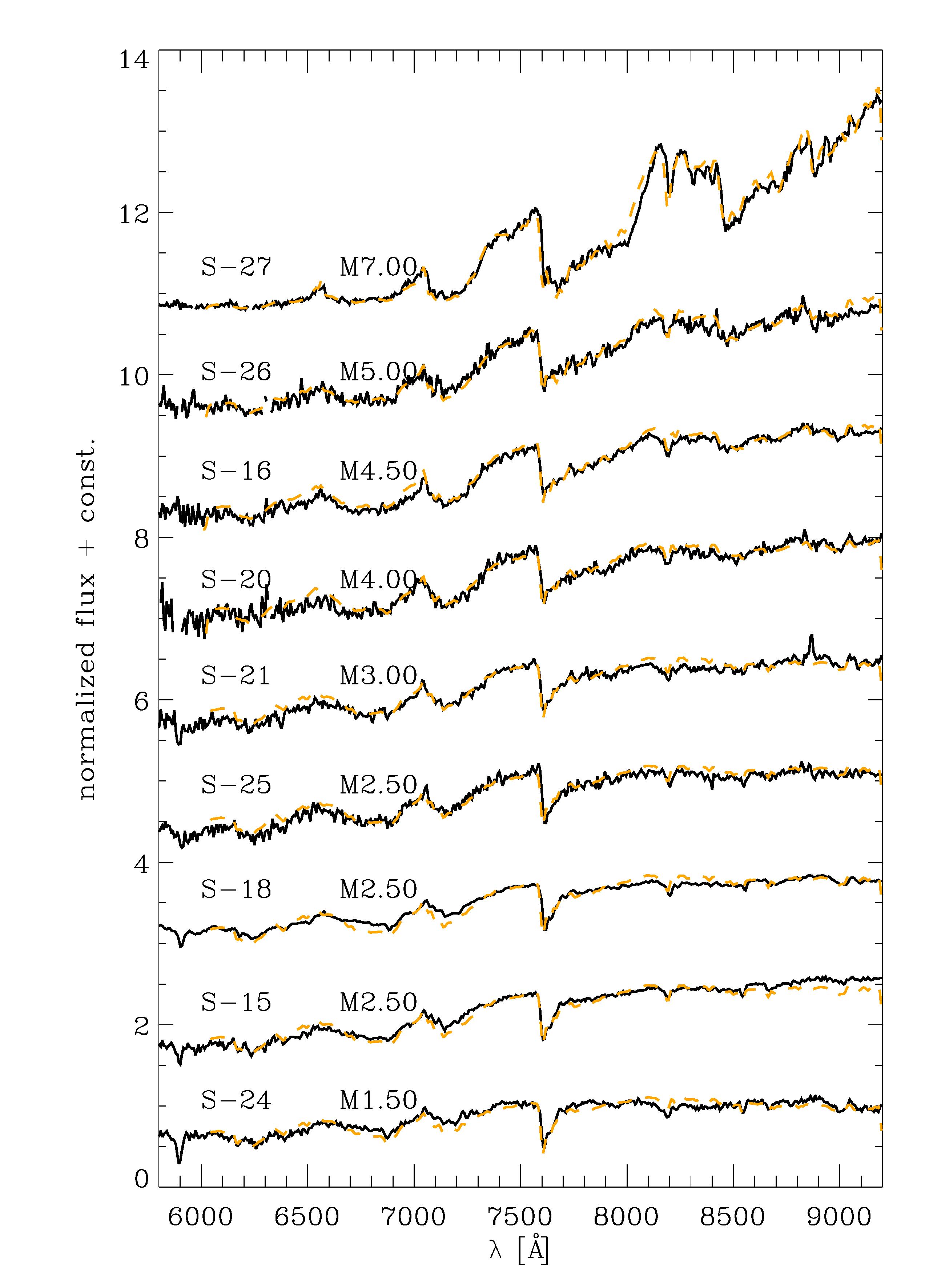}}
\caption{Spectra of the objects identified as field dwarfs (Table~\ref{T_spec}; black), along with the best-fit spectra (orange dashed line).
The spectral resolution of the fit templates has been reduced 
in order to match the resolution of our spectra. Each spectrum has been corrected
for extinction (see Table~\ref{T_spec}), but not for the atmospheric absorption.
}
\label{spec_dwarfs}
\end{figure}

\begin{figure}
\centering
\resizebox{9cm}{!}{\includegraphics{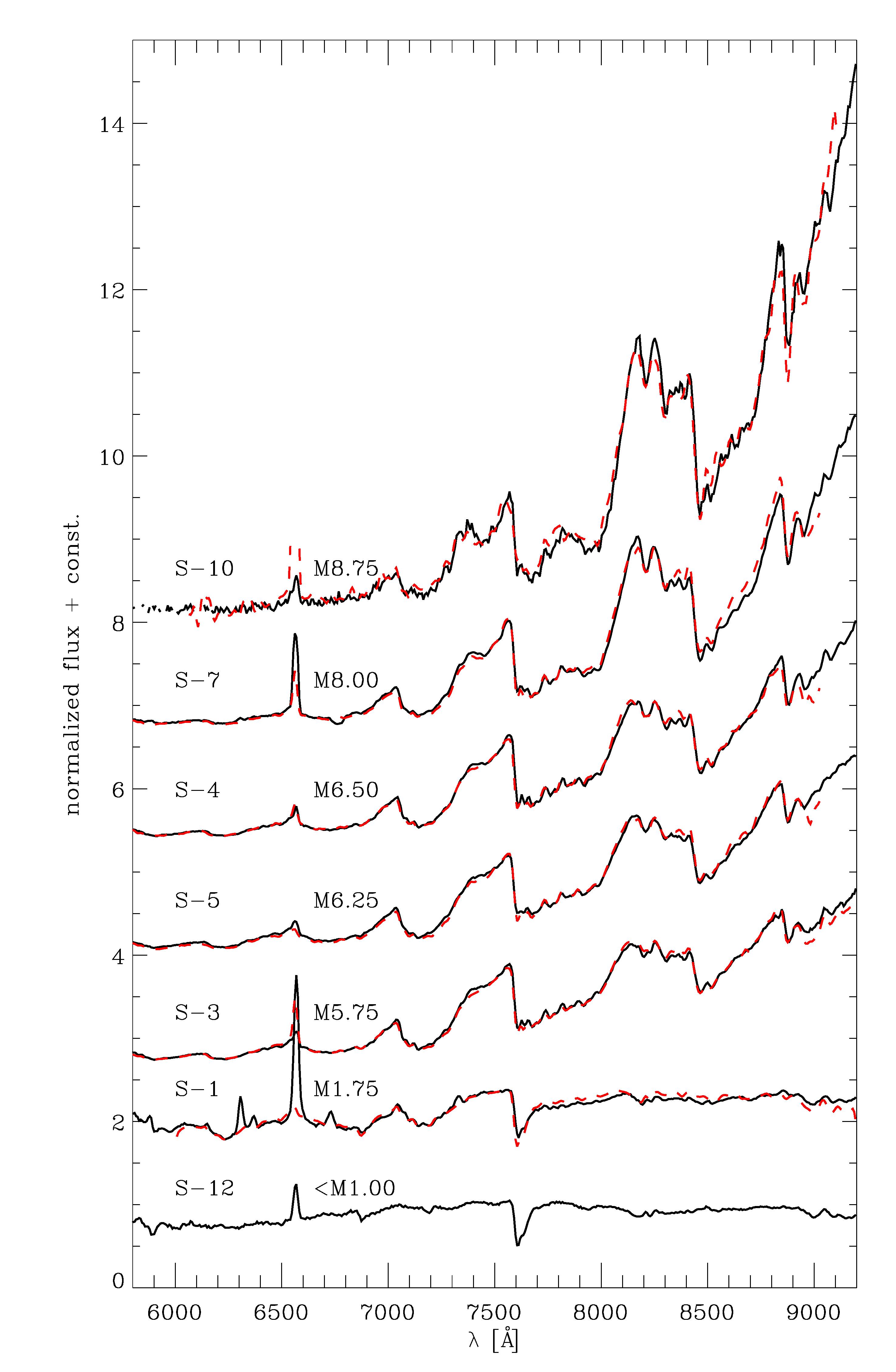}}
\caption{Spectra of the objects identified as probable members of Lupus$\,$3 (Table~\ref{T_spec}; (black), along with the best-fit spectra of members of young star forming regions (red dashed line).
The spectral resolution of the fit templates has been reduced 
in order to match the resolution of our spectra. Each spectrum has been corrected
for extinction (see Table~\ref{T_spec}), but not for the atmospheric absorption.
}
\label{memb_spec_tmpl}
\end{figure}

\begin{figure}
\centering
\resizebox{9cm}{!}{\includegraphics{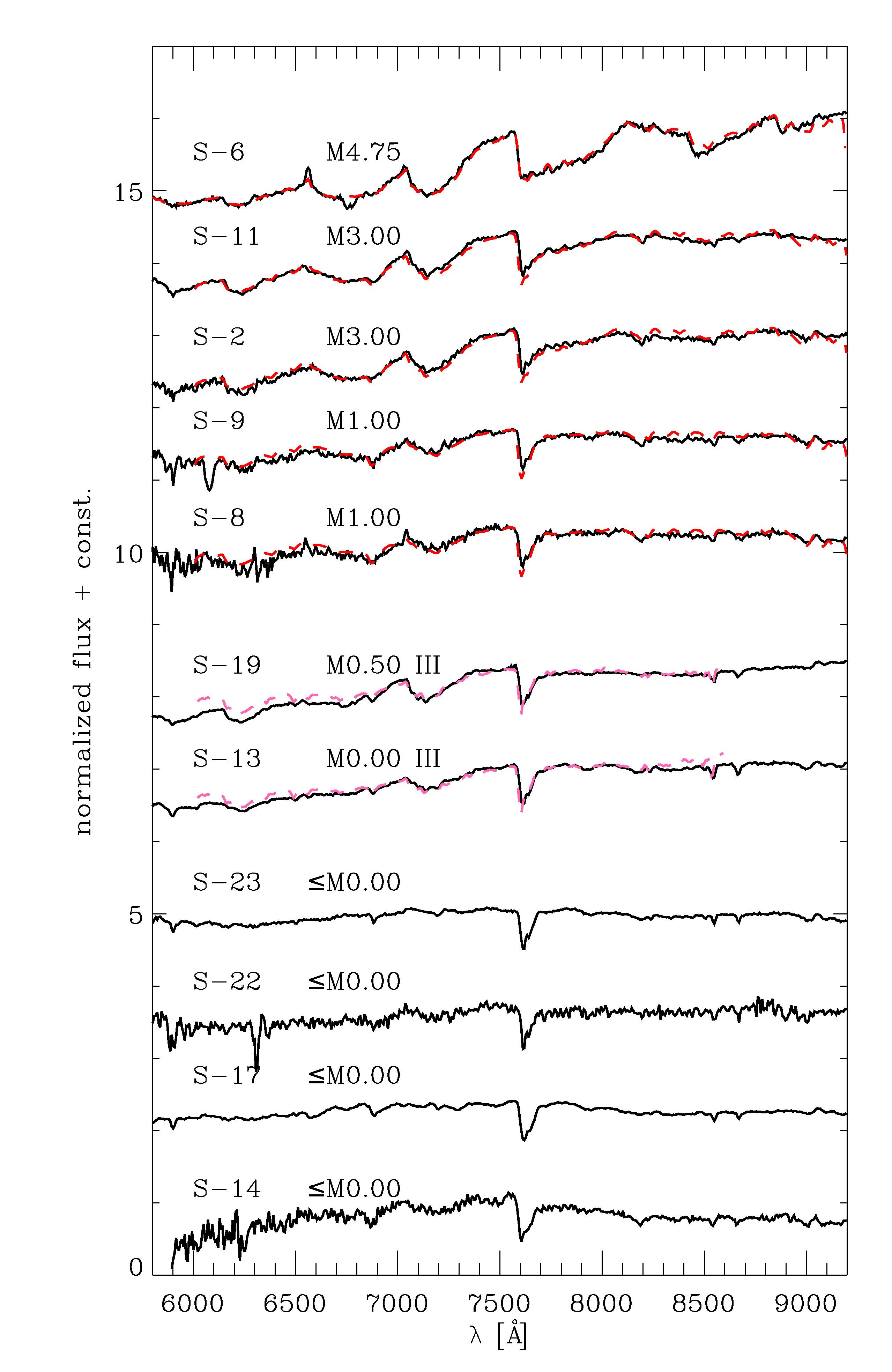}}
\caption{Spectra of the objects identified as giants, or with uncertain classification (Table~\ref{T_spec}; black), along with the 
best-fit spectra (dashed colored line).
The spectral resolution of the fit templates has been reduced 
in order to match the resolution of our spectra. Each spectrum has been corrected
for extinction (see Table~\ref{T_spec}), but not for the atmospheric absorption.
}
\label{spec_giants}
\end{figure}

To determine spectral type and extinction of selected candidate members, we compare
each spectrum to an empirical grid consisting of spectral templates
of field dwarfs, giants, and members of young star forming regions with spectral types M1 to M9. 
A grid of field dwarfs separated by the 0.5 spectral subtype was created by averaging
a number of available spectra at each sub-type. % covering a comparable wavelength range as our spectra 
In case of giants, the grid consist of a collection of available spectra, with a step of 0.5 - 1 spectral subtypes 
\footnote{Spectra of field dwarfs and giants available on \url{http://www.dwarfarchives.org, http://kellecruz.com/M$\_$standards/}, and from \citet{luhman03a, luhman04c}}.
The grid of young objects (1-3 Myr) consists of the spectra of objects in Cha~I, Taurus, and $\eta\,$Cha \citep{luhman03a, luhman04a, luhman04b, luhman04c}, with steps of 0.25-0.5 spectral subtypes.

Spectral type and extinction can be fitted simultaneously. The template spectra
were first smoothed to match the resolution of the VIMOS spectra, and rebinned to the same
wavelength grid. The region around the H$_{\alpha}$ line
was masked. % because the strong emission present in some of our spectra and the templates can affect the
The best fit parameters were determined
by minimizing the test value $\chi$ defined as
\begin{equation}
 \chi=\frac{1}{N} \sum_{i=1}^{N} \frac{(O-T)^2}{T},
\end{equation}
where O is the object spectrum, T the template spectrum, and N the number of data points.

As mentioned before, spectral classification becomes less
reliable at low resolution for objects of spectral type M3 and earlier. It is therefore not surprising that at these early spectral types, 
the fitting scheme gives similar results for different spectral classes. Some level of degeneracy with respect 
to spectral classes is also present at later spectral types, but only in a few cases. It is therefore clear that the final assessment of class (i.e. membership)
cannot be based on this fitting scheme only. However, there is much less degeneracy present with respect to the spectral subtype and A$_V$, also for the objects
with degenerate spectral class. Typically, several best-fit results cluster within $\pm1$ spectral subtype and $\pm1\,$mag in A$_V$, or less, from the best-fit value. 
We therefore adopt these values as typical uncertainties of the fitting procedure.

%The best fit values of $\chi$ are found to be between 0.002 and 0.03. 
%We also inspect each best fit visually, in order to confirm that the spectral
%class returned by the routine is indeed in agreement with our membership criteria described in
%the previous section.
%This is not the case only in a few cases for the objects with spectral types earlier than M3, where the typing becomes less
%reliable at low resolution. 
%These objects will be discussed in more details in the next sections. 
To determine membership status of an object, several best fits at each spectral class are inspected visually, in order to check
for the presence or shape of the spectral features discussed in the previous section. 

Of the initial 27 spectra we started with, 9 objects are classified as field dwarfs because of 
the deep Na~I absorption and the lack of H$_{\alpha}$ emission (see Figure~\ref{spec_dwarfs}).
Two spectra are best represented by spectra of early-M giants, while
four objects have M-dwarf features that are less pronounced than in M1 dwarfs, i.e. are probably
of earlier spectral type (bottom 6 spectra in Figure~\ref{spec_giants}).
The spectra of the latter 4 objects (SONYC-Lup3-14, 17, 22, and 23) all show
evidence for CaII triplet in absorption. The spectra of SONYC-Lup3-17 and 23 also have very weak or non-existent
Na~I absorption, thus are probably background giants. The spectra of SONYC-Lup3-14 and 22 have lower S/N, which poses a difficulty in 
final classification. The depth of the Na~I in the spectrum of SONYC-Lup3-14 hints to a high-gravity atmosphere, but for the sake of caution, we prefer to keep the
membership status of both objects as uncertain.

%The existence of giants in our sample is not surprising, as
%some contamination by background cool giants is expected due to the galactic latitude of Lupus$\,$3 (b=$9.5^{o}$, see discussion in \citealt{comeron09}).

Seven spectra are best fit by young dwarf templates with no degeneracy in spectral class (SONYC-Lup3-1, 3, 4, 5, 6, 7, and 10).
The spectrum of the object SONYC-Lup3-10 is best matched with a M8.75 young dwarf template, that was derived by averaging the spectra
of Taurus members KPNO09 (M8.5) and KPNO12 (M9) from \citet{luhman04c}. 
All of these 7 objects also show H$_{\alpha}$ emission, and appear to have weak Na~I absorption, suggesting some kind of 
low-gravity atmosphere. All 7 objects are also proper motion candidates. 
%11 spectra are best fit by young dwarf templates (named SONYC-Lup3-1 through 11).
SONYC-Lup3-12 exhibits strong H$_{\alpha}$ emission, but its spectrum is outside our fitting grid (earlier than M1). 
The shape of the Na~I feature reveals a low-gravity atmosphere, while the lack of the Ca~II triplet in absorption make
it less probable to be a giant. 
Furthermore, its proper motion argues in favor of its membership in Lupus, and therefore we include SONYC-Lup3-12 in the list of
candidate members.
The spectra of these 8 best candidates for membership in Lupus are shown in Figure~\ref{memb_spec_tmpl}, with exception of SONYC-Lup3-6 which is, for the
reasons that will become apparent in the next sections, classified as uncertain and shown in Figure~\ref{spec_giants}.
 
%At this stage, we do not yet discard any of these 12 objects from our candidate member list, but note
The nature of the remaining 4 objects of early M spectral type (SONYC-Lup3-2, 8, 9 and 11) remains uncertain at this stage. In Figure~\ref{spec_giants} we show their spectra together with the
best-fit young dwarf templates, but we note that other classes might give equally satisfactory fit. The CaH and Na~I regions for all 4 objects fit better in the dwarf
class, rather than giants. The CaII triplet absorption is present, which indicates that none of these objects is young.  
%This issue will be discussed further in the next sections.
%The spectra of the 12 candidates are shown in Figure~\ref{memb_spec_tmpl} (7 spectra), 
%and Figure~\ref{spec_giants} (top 5 spectra). 
%The division is based on arguments presented in following sections.

\subsection{Model fitting}
\label{model_fit}
In order to derive effective temperatures for the SONYC objects, we perform 
the spectral model fitting using the BT-Settl \citep{allard11} and AMES-Dusty \citep{allard01} models. 
Model spectra are smoothed by a boxcar function beforehand, 
to match the wavelength step of the VIMOS spectra.
The three main parameters that affect the spectral shape are the effective temperature ($T_{\mathrm{eff}}$), 
gravity (log$\,g\,$), and interstellar extinction ($A_V$).
Fitting the models to our low-resolution spectra does not allow an unambiguous determination of all the three 
parameters at the same time. We therefore keep log$\,g\,$ constant for each object,
and assume the value of 3.5 (or 4 in case of AMES-Dusty models above 3000~K) for the objects identified as probable young dwarfs,
5 for the field dwarfs, and 3 for the giants (in this case only BT-Settl models are used). For all the
spectra classified as uncertain, we use the intermediate value of log$\,g\,$, i.e. the same as for young dwarfs.
%But, since we are only interested in the objects that were identified as
%members of Lupus$\,$3 in the previous section, we keep log$\,g\,$ constant at the value 3.5, or 4 in case 
%of AMES-Dusty models above 3000~K.
The fitting procedure of 
$T_{\mathrm{eff}}$ and $A_V$ is identical to the one described in \citet{muzic11}.
$T_{\mathrm{eff}}$ is varied between 2000$\,$K and 4500$\,$K in steps of 100$\,$K, and $A_V$ 
from 0 to 10, in steps of 0.5 mag.
To de-redden our spectra, 
we apply the extinction law from \citet{cardelli89}, assuming $R_V\,$=$\,$4.

% Notes:
% some extra checks of suspicious objects
% 67 young - field difficult, but maybe a bite better young. chi2 says young
% 217 fainter has H alpha, but giant for at M5.0 and Av=0.5 fits pretty good
% 10108 field
% 10270 most likely giant
% 10060 likely giant (chi2 agrees)
% 10290 field
% 253 P87 might be young!

%Note on Jan 11: All this fitting with models and by-eye looking on the NaI gravity feature is
% not very efficient, because by quick comparisn to a grid of field dwarfs, giants and Lambda Ori objects,
% I see that it is very easy to distinguish the three categories from the overall shape of the spectra
% plus concentrating on some particular features
% conclusion: do the spectral fitting as a main thins, and then include models to show how they actually fit.

%In Figure~\ref{fld_spec} we show the 7800 - 8600 \AA~ region of the confirmed object \#47, along with the examples
%of four of the objects identified as members of the field population.
The spectra of objects in Table~\ref{T_spec} with the best-fit atmosphere models
are shown in the appendix (Figures~\ref{memb_spec_BT} and \ref{memb_spec_AD}), 
and the results of our spectral fitting procedure for these objects 
are given in Table~\ref{T_models}. 
%We show only the objects with the best-fit
%$T_{\mathrm{eff}}$ at or below 3800$\,$K 
%because the objects found at the edge of the fitting grid could be even warmer.
The uncertainty is 100$\,$K for the
derived effective temperature, and 0.5$\,$mag 
for the extinction, 
which reflects the spacing of the grid. 

%\begin{figure}
%\centering
%\resizebox{5cm}{!}{\includegraphics{NaI_fld.jpg}}
%\caption{Region around the Na~I absorption feature for the Lupus$\,$3 member \#47 (bottom red spectrum), compared to the objects discarded as
%background field objects. We note that the latter objects do not show any signature of H$_{\alpha}$ emission.  
%}
%\label{fld_spec}
%\end{figure}

\begin{deluxetable}{llcccccccccccll}
\tabletypesize{\scriptsize}
\rotate %allow in one-col version!
\tablecaption{Photometry and spectral types for the objects of spectral type M, or slightly earlier, identified towards Lupus$\,$3.}
\tablewidth{0pt}
\tablehead{\colhead{ID} & 
	   \colhead{$\alpha$(J2000)} & 
	   \colhead{$\delta$(J2000)} & 
	   \colhead{$i$} & 
	   \colhead{$J$} & 
	   \colhead{$K$} & 
	   \colhead{A$_{V}^{phot}$} & 
	   \colhead{SpT} & 
	   \colhead{A$_{V}^{SpT}$}	&
	   \colhead{$\mu_{\alpha}cos\delta$} &
	   \colhead{$\mu_{\delta}$}	&
	   \colhead{$\mu-$} &
	   \colhead{EW(H$\alpha$)} &
	   \colhead{membership} &
 	   \colhead{comments} \\
  &	 	&     	       & (mag) &  (mag) &  (mag) & (mag) & & (mag) &(mas\,yr$^{-1}$) & (mas\,yr$^{-1}$) & cand?\tablenotemark{4} & (\AA)}
\tablecolumns{15}
\startdata 
    S-1  & 16:07:08.55 &  -39:14:07.8 & 17.3 & 14.6 & 11.5 &  11.5 & M$\,$1.75 & 3.0 & -4.2$\pm$5.2 & -21.2$\pm$4.7 	& y	& -77.6	 	& Lupus$\,$3 & M08\\				%47
    S-2  & 16:07:33.82 &  -39:03:26.5 & 19.4 & 16.2 & 14.1 &  6.0  & M$\,$3.00 & 5.0 &  0.4$\pm$7.1 & -16.3$\pm$5.9 	& y	& -4.9  	 	& uncertain  &		  \\ 			%67
    S-3  & 16:07:55.18 &  -39:06:04.0 & 16.2 & 13.5 & 11.4 &  5.9  & M$\,$5.75 & 2.0 &-12.4$\pm$7.1  & -22.8 $\pm$5.9 	& y	& -37.2	 	& Lupus$\,$3 & M6.5\tablenotemark{d}, C09, G06 \\	%157
    S-4  & 16:08:04.76 &  -39:04:49.6 & 17.0 & 13.9 & 12.4 &  2.9  & M$\,$6.50 & 1.0 & -8.1$\pm$7.1 & -22.6 $\pm$5.9 	& y	& -44.4		& Lupus$\,$3 & M7\tablenotemark{d}, C09 	  \\	%176
    S-5  & 16:08:16.03 &  -39:03:04.5 & 15.5 & 12.5 & 11.1 &  2.5  & M$\,$6.25 & 1.0 & -9.3$\pm$7.1 & -19.9 $\pm$5.9	& y 	& -37.5	& Lupus$\,$3 & Par-Lup3-1; M7.5\tablenotemark{a}, M5.5\tablenotemark{b} \\		%201
    S-6  & 16:08:59.52 &  -38:56:20.8 & 19.1 & 17.4 & 16.7 &  0.0  & M$\,$4.75 & 0.0 & -6.5$\pm$7.1  & -19.4$\pm$5.9 	& y	& -18.6	& uncertain & 		\\			%217_2
    S-7  & 16:08:59.53 &  -38:56:27.8 & 17.1 & 13.9 & 12.7 &  1.3  & M$\,$8.00 & 0.5 & -8.8$\pm$7.1  & -19.6 $\pm$5.9  	& y 	& -185.7		& Lupus$\,$3 & M8\tablenotemark{c}, G06 \\		%217_1
    S-8  & 16:10:18.13 &  -39:10:33.6 & 19.8 & 16.2 & 13.8 &  7.3  & M$\,$1.00 & 6.5 &-16.2$\pm$5.3 & -9.8$\pm$5.0 	& y	& -4.6		& uncertain & \\					%243
    S-9  & 16:11:26.34 &  -39:02:17.1 & 19.3 & 16.2 & 14.0 &  6.4  & M$\,$1.00 & 5.5 &-14.6$\pm$5.3 & -11.3 $\pm$5.0 	& y 	& \nodata		& uncertain  &  \\				%337
    S-10 & 16:09:13.43 &  -38:58:04.9 & 19.5 & 16.1 & 14.9 &  1.5  & M$\,$8.75 & 0.0 &-13.8$\pm$7.1  & -20.9 $\pm$5.9	& y 	& -48.8	& Lupus$\,$3 &\\					%10610
    S-11 & 16:12:05.91 &  -39:04:07.5 & 18.7 & 15.7 & 14.4 &  1.5  & M$\,$3.00 & 3.0 &-24.7$\pm$5.3 & -20.5 $\pm$5.0	& y	& -5.4 	& uncertain  &\\					 	  %253		
    S-12 & 16:11:18.47 &  -39:02:58.2 & 18.0 & 15.1 & 13.3 &  4.2  & $<$M$\,$1 & \nodata   & -6.1$\pm$5.3 & -8.5 $\pm$5.0 	& y 	& -12.1		& Lupus$\,$3 & \\	%329				 	  %253		
    S-13 & 16:07:36.48 &  -39:04:53.8 & 15.7 & 12.9 & 10.6 &  7.1  & M$\,$0.00 & 3.5 & -3.4$\pm$7.1  & 1.1$\pm$5.9 	& n	&\nodata	& III\tablenotemark{2} &  \\
    S-14 & 16:08:01.42 &  -39:10:16.1 & 19.9 & 16.4 & 14.0 &  7.7  & $<$M$\,$1 & \nodata   & -8.4$\pm$7.1  & -12.7$\pm$5.9	& y	& \nodata	& uncertain & \\
    S-15 & 16:07:52.46 &  -39:09:48.1 & 19.0 & 15.6 & 13.2 &  7.6  & M$\,$2.50 & 4.7 &-29.8$\pm$7.1  & -23.1 $\pm$5.9 	& y	&\nodata	& V\tablenotemark{3} & \\
    S-16 & 16:10:14.95 &  -39:10:16.2 & 19.9 & 16.4 & 14.3 &  5.7  & M$\,$4.50 & 4.7 &  5.9$\pm$5.3 & 24.0 $\pm$5.0 	& n	& \nodata	& V & \\
    S-17 & 16:11:56.11 &  -39:08:23.8 & 15.3 & 12.8 & 11.1 &  4.2  & $<$M$\,$1 & \nodata   & -0.3$\pm$5.3 & 11.1 $\pm$5.0 	& n	& \nodata	& III &  \\
    S-18 & 16:11:09.03 &  -39:06:22.7 & 16.8 & 14.3 & 12.5 &  4.5  & M$\,$2.50 & 3.0 &-11.4$\pm$5.3 & -21.7 $\pm$5.0 	& y 	& \nodata	& V &  \\
    S-19 & 16:11:39.24 &  -39:06:32.8 & 15.4 & 12.5 & 10.4 &  6.0  & M$\,$0.50 & 3.0 & -5.7$\pm$5.3 & -5.4 $\pm$5.0 	& y	& \nodata	& III &  \\
    S-20 & 16:11:36.20 &  -39:02:36.9 & 20.4 & 16.8 & 14.6 &  6.9  & M$\,$4.00 & 5.3 &  3.9$\pm$5.3 & 17.1 $\pm$5.0 	& n	& \nodata	&  V & \\
    S-21 & 16:11:33.34 &  -38:59:44.3 & 19.9 & 16.5 & 14.3 &  4.9  & M$\,$3.00 & 4.5 & 7.8 $\pm$5.3 & 1.1 $\pm$5.0    	& n	& \nodata	& V & \\
    S-22 & 16:10:47.49 &  -38:57:12.4 & 20.4 & 17.1 & 16.2 &  0.0  & $<$M$\,$1 & \nodata   &  \nodata	    &  \nodata			& \nodata	& \nodata	& uncertain  & \\
    S-23 & 16:10:45.40 &  -38:54:54.9 & 16.1 & 14.3 & 12.9 &  2.1  & $<$M$\,$1 & \nodata  & -7.9$\pm$5.3 & -1.9 $\pm$5.0 	& n	& \nodata	& III  & M08, C09 \\
    S-24 & 16:07:09.06 &  -39:01:07.6 & 19.8 & 18.1 & 16.6 &  2.8  & M$\,$1.50 & 1.5 &  7.4$\pm$5.2 & 3.6$\pm$4.7 	& n	& \nodata	& V & \\
    S-25 & 16:12:04.96 &  -39:06:11.4 & 19.9 & 17.9 & 16.0 &  5.1  & M$\,$2.50 & 3.0 & 1.0 $\pm$5.3 & 2.7 $\pm$5.0 	& n	& \nodata	&  V & \\
    S-26 & 16:10:56.40 &  -39:04:31.7 & 19.1 & 17.3 & 15.7 &  3.4  & M$\,$5.0  & 2.5 & 35.4$\pm$5.3 & 0.9 $\pm$5.0 	& n	& \nodata	& V & \\
    S-27 & 16:11:15.08 &  -39:07:15.0 & 18.1 & 16.4 & 14.7 &  3.6  & M$\,$7.0  & 0.0 &  \nodata&  \nodata& \nodata&\nodata & V & \\
  % S-10 & Sz113    16:08:57.79 &  -39:02:22.6 &   --- & 12.3 & 10.9 &  2.1  & M$\,$3.00 & 2.0 &  M6\tablefootmark{a}, M08\\		%607
  % S-10 & 	    16:06:58.70 &  -39:04:05.0 & 15.2 & 13.0 & 12.0 &  0.1  & M$\,$5.25 & 0.0 & C09 \\				%2038	
  % S-12 & Sz95     16:07:52.33 &  -38:58:06.1 & ---  & 11.1 & 10.4 &  0.0  & M$\,$3.25 & 0.0 & M1.5\tablefootmark{c}, M08, G06, C09\\	%10659	
\enddata
\label{T_spec}
\tablenotetext{1}{The IDs SONYC-Lup3-X are abbreviated with S-X;}
\tablenotetext{2}{background giant;}
\tablenotetext{3}{field dwarf;}
\tablenotetext{4}{proper motion candidate;}
\tablenotetext{5}{{\bf C09}- candidate member in \citet{comeron09}, Table~9;
{\bf M08} - IRAC YSO candidate from \citet{merin08}; 
{\bf G06} - X-ray source from \citet{gondoin06}.}
\tablerefs{(a) \citet{comeron03}; (b) \citet{mortier11}; (c) \citet{allen07}; (d) \citet{comeron13}}
\end{deluxetable}

\begin{deluxetable}{lcccccc}
\tabletypesize{\scriptsize}
\tablecaption{Effective temperature and extinction derived from models, for the objects of spectral type M (or slightly earlier) towards Lupus$\,$3.}
\tablewidth{0pt}
\tablehead{\colhead{ID} & 
	   \colhead{$T_{\mathrm{eff}}^{A-D}$} & 
	   \colhead{A$_{V}^{A-D}$} & 
	   \colhead{$T_{\mathrm{eff}}^{BT-S}$} & 
	   \colhead{A$_{V}^{BT-S}$} & 
	   \colhead{$T_{\mathrm{eff}}^{C}$}  &
	   \colhead{A$_{V}^{C}$} \\ 
&	 (K) & (mag) & (K) & (mag)
}
\startdata
       S-1  & 3900 & 3.0 & 3800 & 3.0 & 4400\tablenotemark{a} &	7.1	\\ 	%47
       S-2  & 3600 & 6.0 & 3500 & 5.0 &  \nodata&	\nodata			\\ 	%67
       S-3  & 3100 & 4.0 & 3000 & 3.0 & 2950\tablenotemark{b}  & 2.3 	\\	%157
       S-4  & 3000 & 4.0 & 3000 & 3.5 & 2900\tablenotemark{b} & 3.5	\\      %176
       S-5  & 3000 & 3.5 & 2900 & 2.0 & 2800\tablenotemark{a} & 2.0	\\	%201
       S-6  & 3100 & 0.5 & 3200 & 0.0 & \nodata&  \nodata				\\	%217_2
       S-7  & 2900 & 4.5 & 2800 & 3.0 & 2600\tablenotemark{a} & 1.5			\\	%217_1
       S-8  & 3900 & 6.5 & 3800 & 6.5 & \nodata& \nodata				\\	%243
       S-9  & 3900 & 5.5 & 3800 & 5.5 & \nodata&  \nodata				\\	%337		
       S-10 & 2600 & 4.0 & 2700 & 4.0 & \nodata& \nodata				\\	%10610
       S-11 & 3700 & 4.0 & 3500 & 3.5 & \nodata&  \nodata				\\	%253	
       S-12 & $>$3900 &  \nodata & 4200 & 4.5 & \nodata&  \nodata				\\	%329
       S-13 &  \nodata    &  \nodata   & 3900 & 4.5 & \nodata& 	\nodata			\\
       S-14 & 3900 & 6.0 & 3800 & 6.5    &\nodata & \nodata\\
       S-15 & 3700 & 6.5 & 3600 & 6.0    &\nodata & \nodata\\
       S-16 & 3100 & 5.5 & 3000 & 4.0    & \nodata& \nodata\\
       S-17 & \nodata	   &  \nodata   & 4400 & 5.0    & \nodata& \nodata\\
       S-18 & 3800 & 4.5 & 3600 & 3.5    & \nodata&\nodata \\
       S-19 &  \nodata    &  \nodata   & 3900 & 4.0    & \nodata& \nodata\\
       S-20 & 3400 & 6.5 & 3200 & 5.5    &\nodata & \nodata\\
       S-21 & 3400 & 5.0 & 3300 & 4.5    & \nodata& \nodata\\
       S-22 & $>$4000 & \nodata & 4100 & 2.5    & \nodata& \nodata\\
       S-23 &  \nodata    & \nodata & 4200 & 2.5    & 4400\tablenotemark{a} & 2.7 \\
       S-24 & 3900 & 2.0 & 3800 & 2.0    & \nodata& \nodata\\
       S-25 & 3400 & 3.5 & 3300 & 3.0    & \nodata&\nodata \\
       S-26 & 3100 & 4.0 & 3000 & 2.5    & \nodata& \nodata\\
       S-27 & 2700 & 2.5 & 2800 & 1.5    & \nodata& \nodata\\
\enddata		
\label{T_models}
\tablecomments{{\bf A-D} - AMES-Dusty \citep{allard01};{\bf BT-S} - BT-Settl \citep{allard11};
 {\bf C} - parameters derived by \citet[a]{comeron09} and \citet[b]{comeron13}.
}
\end{deluxetable}

\subsection{MIR excess}

From the list of potential members selected from the IRAC data, we obtained spectra of 11 objects. 
Three of them are confirmed as VLM members of Lupus$\,$3 (SONYC-Lup3-1, SONYC-Lup3-7 and SONYC-Lup3-10). Of the confirmed objects, all
were originally selected from the $iJ$ selection. SONYC-Lup3-1 satisfies the YSO criteria from \citet{merin08}.
The fact that these three candidates exhibit disk excess further strengthens their membership in Lupus$\,$3.

There is only one object in the $Spitzer$ photometric candidate selection (Figure~\ref{MIR_Fig}) that appears as candidate YSO in \citet{merin08}, but it was not confirmed in our work. Judged from the spectrum, 
this object (c2d source number 120; $\alpha=$16:10:45.40, $\delta=$-38:54:54.9) probably belongs to the spectral type K, i.e.
it is too early for our spectral typing scheme to work. %ID=10628
Therefore, while this object does not fall into the class of VLMOs that are the subject of this paper, we do not rule it out as a potential member. 
The remaining spectra also look earlier than M1, and do not show any evidence of H$_{\alpha}$ emission, and are therefore eliminated in
our search for very-low-mass objects.

\subsection{Hertzsprung-Russell Diagram}

\begin{figure}
\centering
\resizebox{8.8cm}{!}{\includegraphics{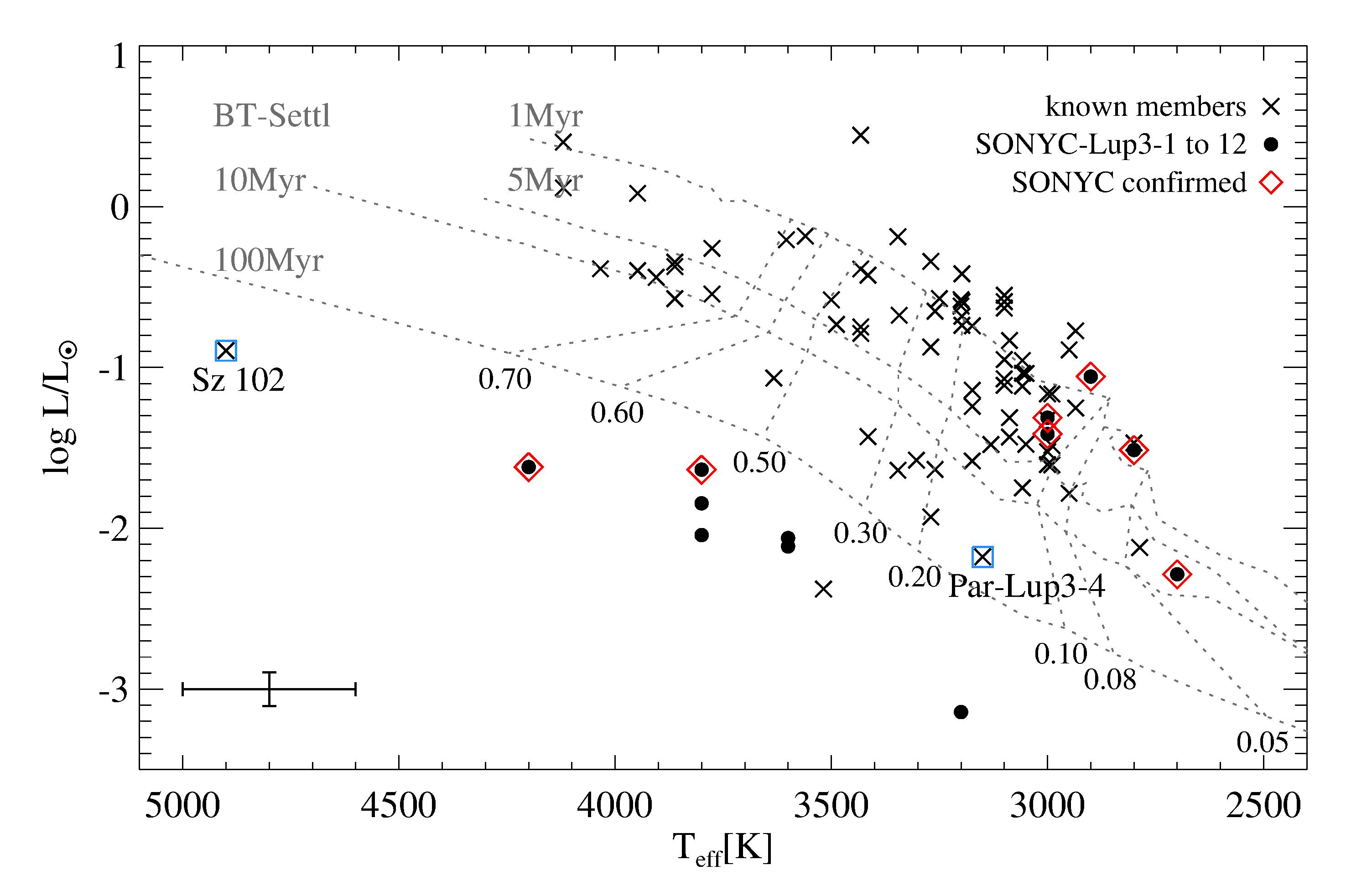}}
\caption{H-R diagram for the Lupus$\,$3 members with existing $T_{\mathrm{eff}}$ 
or spectral type from the literature (crosses). Filled circles mark the sources SONYC-Lup3-1 to -12.
%that were selected as candidate members based on spectral template fitting. 
Red diamonds mark the final subset of confirmed members of Lupus$\,$3. The two well known examples
of underluminous, emission line YSOs are labeled and marked with blue squares. The dashed lines
show BT-Settl theoretical tracks for several ages spanning 1 and 100 Myrs. 
}
\label{HRD}
\end{figure}

In Figure~\ref{HRD} we show the Hertzsprung-Russell diagram, for the list of spectroscopically confirmed members shown in Table~\ref{T_census}.
%with existing estimates of $T_{\mathrm{eff}}$
%or spectral type from the literature shown in Table~\ref{T_census}. 
%We use the absolute $J$-band magnitude
%as a proxy for luminosity, to avoid additional uncertainties from the bolometric correction. 
The extinction was calculated by assuming the intrinsic color $(J-K)_0 = 1$, and extinction law from \citet{cardelli89} with R$_V=4$\footnote{As argued in \citet{scholz09}, the value $(J-K)_0 = 1$ is appropriate for objects of the M spectral type. The only K-type objects shown in the H-R diagram is Sz~102 ($\sim$K2), 
where we adopt a more suitable value of $\sim$0.6 \citep{bessell88}. This change in value shifts the point by $\sim$0.3 dex in the positive y-direction.}. 
%($T_{\mathrm{eff}}\gtrsim$4000$\,$K) this value leads to a slight underestimate in log$L$, as the values between 0.6 and 1.0 are more suitable \citep{bessell88}. 
%The largest influence is in case of Sz~102 ($\sim$K2), where the present point would offset by 0.3 dex in the y-direction. The offset of remaining points (K5-7) is less than the error bar shown in Figure~\ref{HRD}.}. 
The adopted distance is 200 pc.
Where available, we use $T_{\mathrm{eff}}$ estimates available from the literature. 
% with the $T_{\mathrm{eff}}$ estimates available in the literature, we use the  from \citet{comeron09, comeron13, mortier11}. 
For the spectroscopically confirmed members that lack the $T_{\mathrm{eff}}$ information, 
we convert their spectral types to $T_{\mathrm{eff}}$ using the transformation
established in Section~\ref{Teff_SpT}.% (24 sources from \citealt{mortier11} and \citealt{comeron03}). 
%For the K-type stars ($T_{\mathrm{eff}}\gtrsim$4000K) we adopt $T_{\mathrm{eff}}$ as determined by \citet{cohen79}.
We calculate the bolometric correction in the $J-$band (BC$_J$) from the polynomial relation between BC$_J$ and $T_{\mathrm{eff}}$ derived by \citet{pecaut13}
for pre-main sequence stars with ages of 5-30 Myr.
The typical error-bar is shown in the lower left corner. The uncertainty in the $T_{\mathrm{eff}}$ is set to 200 K 
($\sim$200$\,$K for sources from \citealt{comeron09}, $\sim300\,$K for \citealt{mortier11}, and $\sim100\,$K for the rest). 
The error in luminosity takes into account the typical error in the observed $J$ magnitude (0.1 mag), 
distance ($\pm 20$pc), A$_V$ ($\pm 1$ mag), and BC$_J$. We also show the BT-Settl isochrones for ages between 1 Myr and 100 Myr. 
Filled circles mark the 8 objects that were identified as possible members in Section~\ref{spt_fit}, together with the 4 objects with uncertain 
membership status (SONYC-Lup3-1 through 12).
We use $T_{\mathrm{eff}}$ and A$_V$ as derived by comparison with the BT-Settl models. 

The position of sources in H-R diagram can provide additional information about their nature. 
%There is evidence that the 
%ubiquitously observed luminosity spread at a given $T_{\mathrm{eff}}$ in H-R diagrams of young star forming regions is genuine 
%and cannot be attributed to effects of binarity, variability, or accretion \citep{jeffries12a}. 
%Without going into details as to whether this spread indeed reflects a spread in ages or not, 
We can compare the positions of our sources in the H-R diagram, with the positions of the known members of Lupus$\,$3. 
The large fraction of previously confirmed members is located between the 1 and 10 Myr isochrones, with a few also found below the 10 Myr isochrone. 
Among the latter are the two well known cluster members showing significant underluminosity for their spectral types, 
Sz 102, and Par-Lup3-4 \citep{krautter86,comeron03}
\footnote{The spectral type of Sz~102 is not well constrained; for the plot we adopt the $T_{\mathrm{eff}}$ of 4900 K (spectral type K2), according to \citet{mortier11}. 
We note that the associated error in $T_{\mathrm{eff}}$ is $^{+350}_{-840}\,$K, i.e. larger than the typical error-bar shown in the plot.}.

The H-R diagram positions of the five SONYC sources with estimated $T_{\mathrm{eff}}\leq\,$3000\,K agree well with the positions of other Lupus members. 
These sources, SONYC-Lup3-3, 4, 5, 7, and 10, are therefore classified as members of Lupus$\,$3. The remaining sources appear well below the 100 Myr BT-Settl
isochrone.
We compared our photometry with that from 2MASS, to check for variability as a possible cause of the underluminosity for some of these 7 sources. 
While SONYC-Lup3-6 is too faint to be detected by 2MASS, all the rest have photometry of a good quality in at least one of the J- and K-bands. 
SONYC-Lup3-1 shows a difference of 0.7 mag in J, and 0.1 mag in K-band; SONYC-Lup3-12 show a difference of 0.3 mag in the K-band (only an upper limit in J in 2MASS). The remaining sources show no evidence for variability, with differences between the two photometric datasets below 0.1 mag, i.e. comparable to the measurement errors. 
As mentioned in Section~\ref{spt_fit}, SONYC-Lup3-2, 8, 9 and 11 were suspected to be non-members, and Figure~\ref{HRD} confirms it.
The remaining three sources (SONYC-Lup3-1, 6 and 12) deserve special attention and will be discussed below. 
Finally, the lack of the objects with similar spectral types (early-to-mid M) having luminosities that would suggest membership in Lupus
is easily explained by the saturation limit of the $i$-band catalog used for the selection

%Despite a high fraction of disks observed in Lupus (70-80\%; \citealt{merin08}), it is highly unlikely that so many of them should appear to be nearly edge-on.

\subsection{SONYC-Lup3-1}
\label{S-Lup3-1}

\begin{figure*}
\centering
\resizebox{18cm}{!}{\includegraphics{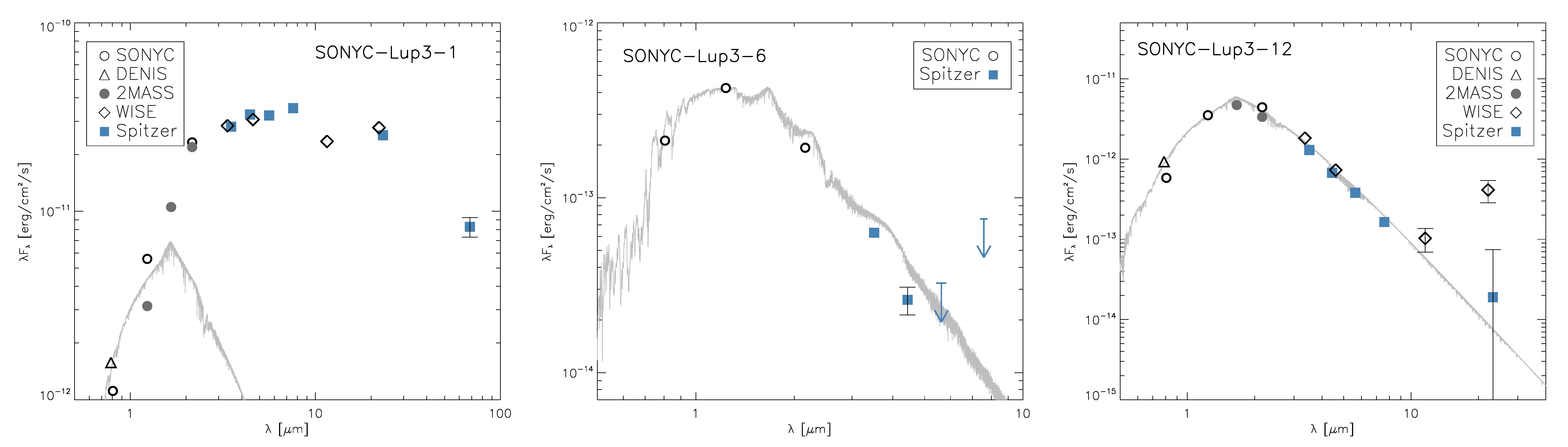}}
\caption{Spectral energy distribution of SONYC-Lup3-1, 6, and 12. 
Photometry comes from this work (open circles), DENIS (triangles), 2MASS (filled grey circles), $Spitzer$ IRAC and MIPS (squares), and WISE (diamonds).  
The uncertainties are comparable to or smaller than the size of the symbols, except for the points with the error bars shown.
}
\label{SED}
\end{figure*}

In addition to being one of the strongest H$_{\alpha}$ emitters among our selected sources (EW(H$_{\alpha}$)=-77.6\AA), 
the spectrum of SONYC-Lup3-1 shows presence of forbidden lines, similar to those previously observed in spectra
of Lupus$\,$3 members Par-Lup3-4, Sz102, and Sz106 \citep{comeron03}. We identify the [OI] lines at 6300\AA~and 6364\AA, [SII] line at 6731\AA~(possibly blended with [SII] line at 6716\AA), and [OII] at 7329\AA. Forbidden lines such as those identified here are commonly observed in actively accreting young stars, and their ratios can be used to estimate physical conditions in the circumstellar medium and mass-loss rates \citep{osterbrock89}. However, as discussed in \citet{comeron03}, forbidden lines in T-Tauri spectra are usually split into separate components sampling different physical conditions, which cannot be distinguished with our low-resolution spectra. Moreover, our spectra do not allow definitive identification
of some lines (e.g. [SII] at 6731 and 6716 \AA). We therefore refrain from physical interpretation of the conditions causing this emission, and leave this
to future studies at higher spectral resolution.

Two sources in Lupus$\,$3 are reported in the literature to exhibit similar properties.
Sz~102 appears underluminous for its spectral type, which was 
was first noted by \citet{hughes94}. \citet{krautter86} found Sz~102 to be associated with a bipolar jet and a Herbig-Haro object. This geometry suggests that we 
might be viewing the star through a nearly edge-on disk \citep{krautter86,hughes94}. \citet{comeron03} report the underluminosity of Par-Lup3-4, and
a jet associated with Par-Lup3-4 was discovered in [SII] images by \citet{fernandez05}.
\citet{comeron03} discuss three possible interpretations for the underluminosity and emission line spectra of the two objects:\\
(a) An edge-on disk blocking a part of the stellar light,   
%While an {\it edge-on disk} can explain the underluminosity of an object, it is unlikely that it can at the same time be responsible for the 
%strong emission lines. In principle, if the lines are formed in a region well detached from the photosphere and not occulted by the disk, they might appear 
%superimposed on a photospheric continuum seen only in scattered light. However, from the consideration of various line ratios, the authors conclude that they most likely form in a vicinity of the central object, and should then also be observed in objects with normal luminosity. Since the strong forbidden lines
%in emission have been only associated with the underluminous objects, the edge-on disk scenario seems to be highly unlikely cause of the both phenomena.\\
(b) embedded Class I sources where the star is seen via the light scattered by the walls of cavities
in the envelope, and 
%with similar emission-line spectra have been observed by \citet{kenyon98} in Taurus. However, none of the underluminous sources in \citet{comeron03} appear extended (optical sizes of $5-10''$ for the Taurus sources), and not all of them show infrared excess.\\
(c) accretion-modified evolution, similar to what is described in \citet{hartmann98} and \citet{baraffe09}.
%A good qualitative description of the phenomenon is provided by the effects of {\it accretion}, which can accelerate the luminosity evolution of a star in a way such that the star appears in the HR-diagram at the same position that %would be occupied by a non-accreting star of the same mass but of older age \citep{hartmann98}.
%However, the magnitude of the observed effect is much higher than the effect estimated by \citet{hartmann98}. \citet{comeron03} speculate that the effect might be more dramatic at masses closer to the BD regime.\\
High-resolution spectra of Par-Lup3-4 \citep{fernandez05} reveal a double-peaked [SII] profile, implying that the low excitation jet is seen at a small angle
with respect to the plane of the sky. The authors argue that at this inclination, only a flared disk could hide a star. \citet{huelamo10}
studied the target SED from the optical to the sub-millimeter regime, and
compared it to a grid of radiative transfer models of circumstellar disks. They find that Par-Lup3-4 is a Class II source, with an edge-on disk that naturally explains its underluminosity.

In Figure~\ref{SED} we show the SED of SONYC-Lup3-1 in the range 0.8 - 70 $\mu$m\footnote{constructed with the help of VOSA \citep{bayo08}}. Open circles mark the optical and NIR photometry from the SONYC campaign, the filled grey
circles show the 2MASS NIR photometry, squares $Spitzer$ MIR data, and diamonds the MIR photometry from WISE. 
%We first note that even the optical-NIR part of the SED
%cannot be adequately fit with any of stellar atmosphere models within a few hundred K around the $T_{\mathrm{eff}}$ derived from the optical spectrum. 
The double-peaked 
shape of the SED with a dip around 10 $\mu$m is characteristic for an edge-on disk, where the NIR peak arises from the scattered light of the star obscured by a disk, 
and the emission at longer wavelengths comes from the dust \citep[e.g.][]{wood02,pontoppidan07,huelamo10}. 
The radiative transfer SED modeling could shed more light on the properties of the putative disk around SONYC-Lup3-1, but since this is
out of the scope of this paper, we refrain from discussing this source further.
To calculate the luminosity of SONYC-Lup3-1 in Fig~\ref{HRD}, we used the A$_V$=3.0$\,$mag derived from fitting a model to the spectrum. The extinction calculated 
from the $J-K$ color is, however, significantly higher (A$_V=11.5$\,mag), and shifts the point in the H-R diagram in the y-direction such that it falls roughly onto the 10 Myr 
isochrone.
We conclude that SONYC-Lup3-1 is probably a member of Lupus$\,$3, with properties similar to Par-Lup3-4 and Sz~102. The MIR excess and 
the double-peaked SED suggest a presence of an edge-on circumstellar disk. 

\subsection{SONYC-Lup3-6}
\label{S-L3-6}
%In the color-magnitude diagram used for the photometric selection (Figure~\ref{IJCMD}), 
%one object is marked as confirmed, but lies outside the
%pre-main-sequence selection box. 
We obtained the spectrum of SONYC-Lup3-6 by chance alignment of the slit that was placed on SONYC-Lup3-7, which belongs to our high-priority candidate list.
SONYC-Lup3-6 is located $\sim 7''$ north of SONYC-Lup3-7.
With the $i-J$ color of 1.8, SONYC-Lup3-6 is too blue to be in our photometric candidate list, i.e. the colors suggest that the object is not a pre-main sequence source.  
In the HR diagram, it falls well below the 100 Myr isochrone.
However, the spectral fitting procedure identifies the object as a young star of the spectral type M4.75. 
It shows low-gravity Na~I signatures, H$_{\alpha}$ emission, and no evidence for Ca~II triplet absorption.
Furthermore, its proper motion satisfies the proper-motion membership criterion. From the EW(H$_{\alpha}$) it is not clear whether SONYC-Lup3-6 undergoes accretion or not, 
as it falls close to the line that separates accretors from non-accretors (yes according to the criterion by \citealt{barrado03}, and no according to \citealt{white&basri03}). 
If indeed a member of Lupus$\,$3, SONYC-Lup3-6 might be a primary of a very wide binary system containing the M8 brown dwarf SONYC-Lup3-7.
The projected separation of this hypothetical system would be $\sim$1400 AU. It would be, however, very difficult to explain 
a M4.75 primary that is several magnitudes fainter than its supposed secondary. We therefore classify the membership of SONYC-Lup3-6 as uncertain.

\subsection{SONYC-Lup3-12}

This object's spectrum suggests a type earlier than M1, but was nevertheless selected in our preliminary analysis based on the obvious H$_{\alpha}$ emission observed
in its spectrum (Figure~\ref{memb_spec_tmpl}). According to \citet{west08}, early M-dwarfs have shorter activity lifetimes when compared to the later M-dwarfs, 
which speaks in favor of SONYC-Lup3-12 being indeed young. However, as in the case of SONYC-Lup3-6, the EW(H$_{\alpha}$) falls right at the border separating accreting objects from non-accretors 
(accretor according to the criterion by \citealt{barrado03}, and non-accretor according to \citealt{white&basri03}).
When compared to the spectra of other objects in our sample
with similar early spectral type (e.g. SONYC-Lup3-17, 22, 23), we notice the lack of the CaII absorption lines, i.e. this object is likely not
to be a giant. 
Based on this evidence, we tentatively include SONYC-Lup3-12 in the final list of the confirmed members.
It shows no MIR excess, and therefore the edge-on disk morphology does not
seem to be a plausible explanation for its underluminosity. 
%There are, on the other hand, other young dwarfs showing
%similar properties (underluminosity, strong H$_{\alpha}$ emission, and no MIR excess), e. g. Sz~113 (member of Lupus$\,$3) and LS-RCrA 1
%\citep{comeron03, barrado04, fernandez05}. -> the statement with MIR excess not true!!! They talk about NIR excess in those papers

%From Aleks: I just checked in the WISE database, and both Sz 113 and LS RCrA-1 have clear excess (W1-W2>0.5 or so, detected in W3 and W4). 
%But S-12 doesn't. S-6 is an interesting case, because the nearest WISE neighbour is 6" away. If our coordinates are correct (the ones in Table 2), I don't know what that means, maybe it is really too faint.

\subsection{Summary of the results}

From the candidate list containing 409 objects in the direction of Lupus$\,$3 cloud, 123 objects were selected for the spectroscopic follow-up. 27 showed
spectral features consistent with spectral type M, or slightly earlier. Fitting the spectra to an empirical grid of spectra of young dwarfs, 
field dwarfs and giants, combined with the proper motion assessment of membership, and the positions of the candidates in the HR diagram, we identify 9 field dwarfs, 4 background giants, and 7 probable members of Lupus$\,$3, among which two show significant underluminosity for their spectral type.
The nature of 7 sources remains uncertain.
In Table~\ref{T_spec} we list the 27 analyzed sources, and give their coordinates, photometry, spectral types, proper motions, EW(H$_{\alpha}$), membership status, and
references for those identified in the literature.

\section{Results and discussion}
\label{discuss}

In this section we use the members of Lupus$\,$3 confirmed in this work to derive a relation between spectral type and $T_{\mathrm{eff}}$, and 
combine the outcome of our survey with the 
results from other works, in order to establish a census of VLM objects in Lupus$\,$3, and discuss the properties of the substellar population.

%As shown in Figure~\ref{memb_spec_tmpl}, 12 objects are confirmed as probable low-mass 
%members of Lupus$\,$3, and are assigned spectral types ranging between M1 and M8.75. Four of 
%the objects (SONYC-Lup3-4, -5, 7 and -10) have spectral types later than M6 and are probably substellar. 
%SONYC-Lup-7 was previously typed as M8 by \citep{allen07} from its 
%optical spectrum, in agreement with the spectral type obtained in this work. 
%SONYC-Lup-10 is the object with the latest spectral type we found (M8.75), and was not reported in previous works.

%The third object with spectral classification from the litearure is  SONYC-Lup3-12 (known as Sz 95). It was classified as M1.5 by \citet{hughes94}, while we find
%a spectral type that is almost two subtypes later (M3.25). \citet{comeron03} note a systematic offset in spectral types for the common objects in the two works, with those 
%from \citet{comeron03} being about 2 spectral subtypes later than those in \citet{hughes94}. The authors argue that the 
%5800 - 7000 \AA interval used in \citet{hughes94}, is probably better suited for the classification of earlier spectra, and less reliable when compared 
%to a wide wavelength range such as ours. The $T_{\mathrm{eff}}$ for this object is in agreement wit the one from \citet{comeron09}.
%One object has a spectral type M5.75, i.e. it is right at the border between stars and BDs. The model fitting gives a $T_{\mathrm{eff}}\sim 3050\,$K, again in agreement with 
%$2900$\pm$200$\,K found in \citet{comeron09}.

\subsection{Effective temperature vs spectral type}
\label{Teff_SpT}

\begin{figure}
\centering
\resizebox{9cm}{!}{\includegraphics{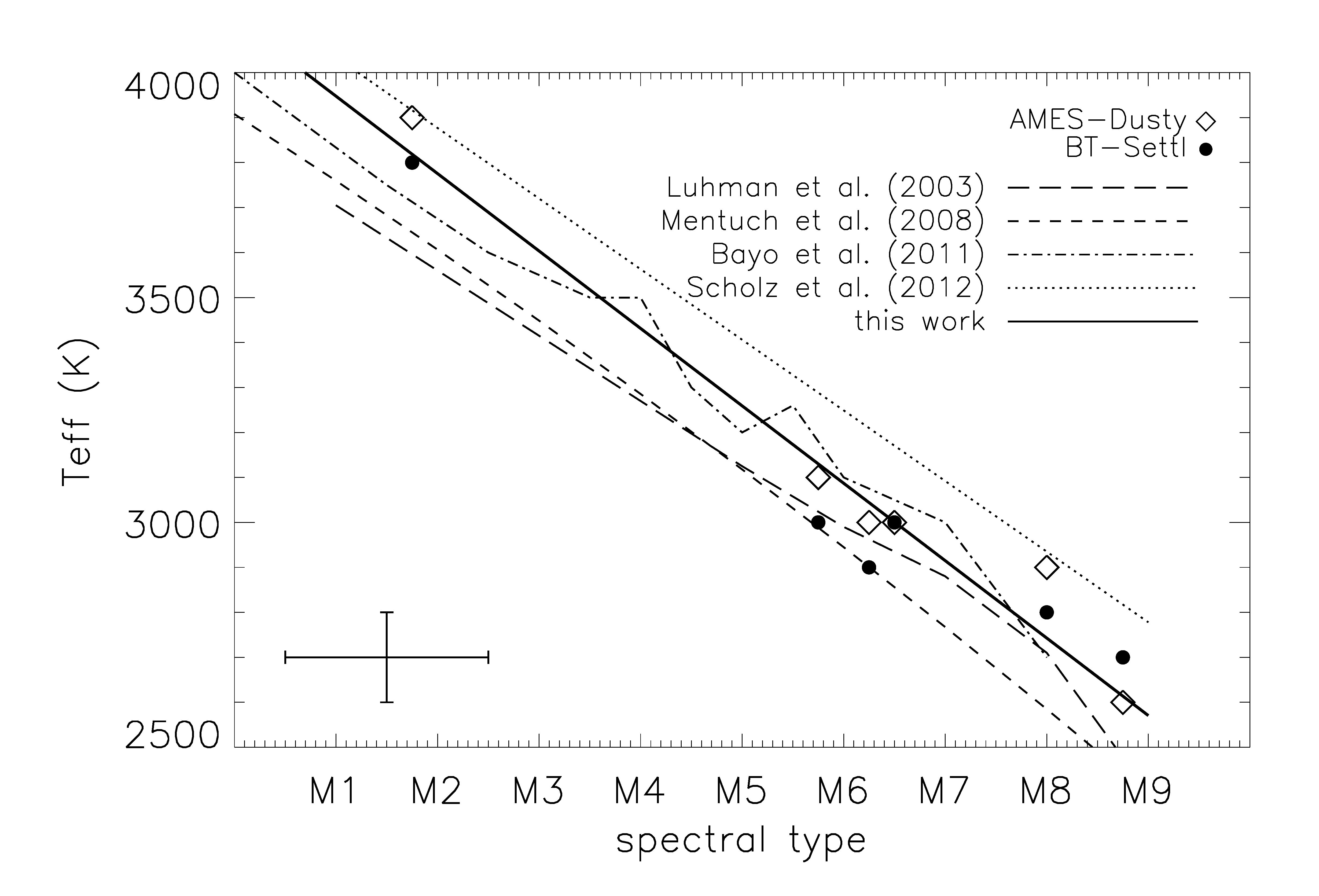}}
\caption{Comparison between the spectral types and effective temperatures for the probable members of Lupus$\,$3 with $T_{\mathrm{eff}}<4000\,$K confirmed in this work. 
Two sets of symbols stand for effective temperature that were determined by fitting different models to our spectra (see Table~\ref{T_models}).
The typical error-bar is shown in the lower left corner of the plot. The solid line is a linear fit to the data points. 
We also show the effective temperature scales by \citet{luhman03b, mentuch08, bayo11,scholz12a}. 
}
\label{spt_teff}
\end{figure}

In Figure~\ref{spt_teff}, we plot the relation between $T_{\mathrm{eff}}$ and the spectral type for the objects classified as members of Lupus$\,$3 and with $T_{\mathrm{eff}}<4000\,$K
(SONYC-Lup3-1, 3, 4, 5, 7, and 10).
The upper limit in $T_{\mathrm{eff}}$ is set because our spectral type fitting scheme does not extend to spectral types earlier than M1. 
Each object listed as member in Table~\ref{T_spec} is represented by two data points, one for the $T_{\mathrm{eff}}$ derived using
BT-Settl models (filled circles), and the other using AMES-Dusty (open diamonds). The solid line
represents a linear fit to all the data points: 
\begin{equation}
T_{\mathrm{eff}} = (4120 \pm 175) - (172 \pm 26) \times \mathrm{SpT}, 
\end{equation}
where SpT corresponds to the M subtype.
A linear least-square fit is performed using the $fitexy$ procedure from \citet{press02}, which takes into account both the uncertainties in spectral type and $T_{\mathrm{eff}}$, shown in the lower left corner of Figure~\ref{spt_teff}. The reduced $\chi^2$ of the fit is $2$, with the $q$ parameter of 0.996\footnote{The $q$-parameter stands for the probability that a correct model would give a value equal or larger than the observed $\chi^2$. It is a scalar with values between 0 and 1, where a small value of $q$ indicates a poor fit.}.
We also show the effective temperature scales available from other works. The long-dashed line shows the scale from \citet{luhman03b}
derived for IC348, while the dashed line represents the scale derived for the nearby young stellar associations by \citet{mentuch08}. 
The latter scale is extrapolated below 3000~K. Dash-dotted line shows the scale derived from the spectroscopic data of Collinder 69 \citep{bayo11}, 
and the dotted line is a scale for NGC1333, derived by using the H-band peak index (HPI) defined in \citet{scholz12a}. This scale was extrapolated to the spectral types earlier than M6, 
since the HPI index is defined only for later spectral types. We note that the scale of \citet{scholz12a} was derived from the near-infrared data, 
while all the rest are based on optical spectroscopy. 

Although the trend seen in our data is generally consistent with all the previous works, some systematic offsets between the classification
schemes is evident. The HPI index seems to give systematically later spectral types for a given $T_{\mathrm{eff}}$ in comparison to other
schemes. The same trend was already reported in the \citet{muzic12}, where we used the HPI to derive spectral types from NIR spectra in highly extincted cluster $\rho$~Oph. 
There we speculated that the extinction might introduce additional uncertainties to spectral type derived from the HPI, because the index was 
defined on a sample of VLMOs in NGC~1333, a cluster with A$_V$ typically below 5. Lupus$\,$3, however, has extinction levels comparable to those of NGC~1333, while we
observe the same trend towards later spectral types. It seems therefore that for HPI the trend persist irrespective of differences in extinction of the samples, and wavelength ranges
used for spectral classification (optical versus NIR).   
A widely-used relation between spectral type and $T_{\mathrm{eff}}$ from \citet{luhman03b} seems to deliver systematically lower $T_{\mathrm{eff}}$ for the same spectral type, 
 compared to the relation derived here, especially for spectral types earlier than M6. 

\subsection{Census of very low-mass members of Lupus$\,$3}
\label{liter}

The most recent summary of the population of the Lupus star forming region appeared in \citet{comeron08}. Since then, however, various studies have identified new low-mass
members, and most importantly, there have been a significant efforts to provide spectroscopic confirmation of photometrically
identified candidates. We therefore think it timely to create a summary of the previous studies, together with results of our work, and provide
a census of the very low-mass content of Lupus$\,$3. In Table~\ref{T_census}, we list coordinates, photometry, spectral type and/or $T_{\mathrm{eff}}$, and alternative 
names for all {\it spectroscopically confirmed} members with the spectral type M0 or later.
%, or $T_{\mathrm{eff}}$ below 3200$\,$K. 
%For each study, we also
%make a comparison with our dataset. Note that we make a cutoff at $i=15.3\,$mag, and thus exclude all the bright stellar members classified in early surveys of the region.
Since we choose to concentrate only on the spectroscopically confirmed members, we do not include the photometric candidates
from the studies such as \citet{nakajima00} and \citet{lopez-marti05}, although we do include the identifiers from the latter study, in case they were 
spectroscopically confirmed in other works.
%some objects to additionally check and maybe include:
%Sz 102 (the one with many emission lines) from C03
%mortier 2011 -> check if some of these are discarded as members

%Krautter et al. (1997) T-Tauri stars in Lupus3: 15 earlier than M, 16 M type
%Comeron 2008 Table 12 summary of all CTTS in Lupus3. total 43. 8 earlier than M0, 2 without classificiation. Rest M.

The starting point for this section is the review paper on the content of the Lupus clouds by {\bf \citet{comeron08}}. Table~11 of the review contains
a list of all classical T Tauri stars in Lupus$\,$3 known to the date of publication. This list includes all the bright members identified and confirmed in studies by 
\citet{the62}, \citet{schwartz77}, \citet{krautter92}, \citet{hughes94}, and \citet{krautter97}, among which are 33 M-type stars.

%\begin{itemize}
%\renewcommand{\labelitemi}{$\bullet$}

{\bf \citet{comeron03}} surveyed a small area ($\sim10' \times 5'$) surrounding the two brightest members HR5999 and HR6000, using slitless
spectroscopy followed by the MOS follow-up using FORS/VLT. They find four M-type members, including Par-Lup3-1 (SONYC-Lup3-5), which the authors label 
a bona-fide BD because of the assigned spectral type of M7.5, and presence of the signatures of youth. 
Our spectral classification assigns it a slightly earlier spectral type (M6.25). 
The difference in the spectral types might arise from the different spectral typing schemes, especially since the typing in \citet{comeron03} was 
done by comparison with the field stars from \citet{kirkpatrick91}, as not many young, low-gravity BDs were known at the time.
The $T_{\mathrm{eff}}$ derived from the model fitting is, on the other hand, in excellent agreement with $2800\pm200$\,K derived from the SED modeling in the 
optical and near-infrared in \citet{comeron09}. %The same object was classified M5.5 in \citet{mortier11}.
%One more object from the list in \citet{comeron03} matches our IJ catalog, but with $i=15.5$ and $i-J=0.8$ falls to the left of the area shown in Figure~\ref{IJCMD}. 
%The object is labeled Sz 102, but was not classified since the spectrum appears featureless, except for the abundant emission lines.

%\item {\bf \citet{lopez-marti05}} identify 22 candidate members from multiband photometry and H$_{\alpha}$ emission. From this list, 7 are found in our $iJ$ catalog, 
%but only one actually matches our candidate list (Lup-831). The object is not found in the proper-motion selected list, and was not observed spectroscopically.

%\item 
{\bf \citet{allen07}} spectroscopically confirmed six candidate members selected from $Spitzer$ MIR photometry, %Two of these objects are found in our $iJ$ catalog, 
one of which was also in our high-priority selection box (SONYC-Lup3-7). \citet{allen07} classify SONYC-Lup3-7 as M8, in agreement with the spectral type obtained in this work. 

%\item 
{\bf \citet{merin08}} selected 124 candidate members in Lupus$\,$3 from the $c2d$ Spitzer IRAC observations. 
Part of this sample (46 objects) was spectroscopically followed-up with FLAMES/VLT, by {\bf \citet{mortier11}}. From the 46 objects, 8
are discarded as non-members because they lie far above the PMS evolutionary tracks. For the three objects found below the tracks, the authors propose
an envelope or a edge-on disk geometry in case of membership in Lupus. Otherwise, these would be sources located in the background. One of these three objects is a well known
underluminous member of Lupus$\,$3 (Sz~102), and an emitter of forbidden lines (see Section~\ref{S-Lup3-1} for more details). 
We choose to
include one of the two remaining objects to the compiled list of members (M-65), since it exhibits MIR excess, H$_{\alpha}$ emission, and spectral type later than M0. The third object is classified as K2 (M-71) and thus not included in the census table.
%who classified 46 of the MIR-excess sources as members of Lupus$\,$3, 
%using the optical spectra obtained with FLAMES/VLT. 

Two of the spectroscopically confirmed objects from \citet{mortier11} are found within our photometric selection box, and for one of them
we also obtained a spectrum. This is SONYC-Lup3-5 (Par-Lup3-1), which Mortier et al. classify as M5.5. 
In this work the same object is classified as M6.25, and by \citet{comeron03} M7.5.
Spectra used in \citet{mortier11} span only the blue part of our spectral range (6550-7150 \AA), which might
be the cause of the different spectral classification, as the molecular features at $\lambda > 7200 $\AA~show more dramatic change with spectral type than the blue portion of the optical spectra. 
%The sample is divided in two groups: candidate members (new objects identified in that work), and PMS-stars (if identified as candidate in other surveys, or previously spectroscopically confirmed). From the PMS-star list, two objects are found in our 
%candidate list. The PMS status of these two objects are deduced from the spectroscopic confirmation of \citet{allen07} (SONYC-Lup3-7), and \citet{lopez-marti05} (Lup-831).
%From the IRAC candidates identified by \citet{merin08}, two are found in our selection box, and the one followed-up by spectroscopy is indeed confirmed as a low-mass pre-main sequence star (SONYC-Lup3-1). 
%The other one was followed up by \citet{mortier11} and also confirmed.

%\item 
{\bf \citet{comeron09}} identified 72 candidate members of Lupus$\,$3 with $T_{\mathrm{eff}}< 3400\,$K. Recently, a spectroscopic follow-up of 
46 out of 72 candidate members by {\bf \citet{comeron13}} revealed that about 50\% of the objects are background giants.
Only 4 objects from the sample of \citet{comeron09} are in the SONYC area, and not affected by saturation and data reduction artifacts in our optical images. 
One of these 4 objects (16:09:20.8, -38:45:10) appears elongated in both MOSAIC and NEWFIRM images, and thus did not end up in our final $iJ$ catalog. 
Its spectrum by \citet{comeron13} shows that it is probably an M-type field dwarf.   
The remaining 3 objects are all found in our high-priority candidate list (photometric and proper motion selection), 
2 of which were observed spectroscopically and indeed confirmed as members (SONYC-Lup3-3 and -4). 
% extended source 242.33667 -38.752778 NEW F8

Recently, {\bf \citet{alcala14}} published the VLT/X-Shooter spectroscopy of several Class II YSOs in Lupus~3, selected from the lists
of \citet{merin08} and \citet{allen07}. Due to the exceptionally wide wavelength range of X-shooter spectra (UV to NIR), and their resolution, we expect that the spectral types derived from these observations should be highly reliable, and are included in the census in Table~\ref{T_census}.

%Merin 34 with I>15.3 in our FOV, 27 matched with $iJ$ cat (some in the gaps), only 5 in the cand list 

\subsection{Substellar population in Lupus$\,$3}
\label{missing}

Based on the survey presented in this work, we can put
limits on the number of VLM sources that are missing in the
current census of YSOs in Lupus$\,$3.
To avoid a possible bias towards the objects with disks, we
 base the following discussion only on the $IJ$ candidate list, and take into account only the objects above the completeness limit of our survey (dotted
line in Figure~\ref{IJCMD}). The completeness limit of $i=20.3$ is equivalent 
to 0.009 - 0.02 \solm\ for Av=0-5, at a distance of 200$\,$pc and age of 1 Myr, according to the BT-Settl models. The limits translate to
0.007 - 0.02 \solm\ in case of AMES-Dusty, and 0.005 - 0.015 \solm\ in case of AMES-COND models \footnote{Note that the version of DUSTY and COND models
with updated opacities, BT-Dusty and BT-COND, do not extend below 0.03 \solm\ at early ages.}.

In the candidate selection box in Figure~\ref{IJCMD} we find 342 candidates above the completeness limit.
Of these, 54 are also proper motion candidates (i.e., belong to the ``$IJ$-pm'' list). We obtained 111 spectra for the photometrically selected candidates, from which 31
are also proper motion candidates. The 7 objects confirmed to be Lupus$\,$3 members in this work are all from the proper-motion selection.
To this we add three previously confirmed members found inside the box (green stars in Figure~\ref{IJCMD}), from which one is a proper-motion candidate.

Our experiment is an example of a random variable, following the binomial probability distribution. Based on the observed number of success counts in our sample, we can determine the maximum-likelihood value for the frequency of substellar objects in the sample, which is basically the ratio between the confirmed objects and those objects with spectra. The 95\% confidence intervals (CI) are calculated using the Clopper-Pearson method, which is suitable for small-number events, and returns conservative CIs compared to other methods \citep{gehrels86, brown01}
The success rate in the ``$IJ$-pm'' sample is therefore $8/32 = 25\substack{+18 \\ -14}\%$, while for the photometric candidates it is $2/82\approx2\substack{+6 \\ -2}\%$.
In total, the estimated number of unconfirmed members in our survey is $(54-32)\times 0.25 + (342-54-82)\times0.02 = 6 + 5 = 11$,
 with the 95\% CI of $\substack{+13 \\ -5}$.
We can repeat the same calculation to get an approximate number of missing BD above the completeness limit. From the SONYC sources, 3 are most probably BDs (all existing spectral classifications 
$\geq\,$M6), while two have spectral types around the substellar border. Among the three sources from other works, one is a star earlier than M0, one 
is classified as M7.5 (J16083304-3855224 in Table~\ref{T_census}), and the remaining one is classified as M6 (M-66). 
The total number of spectroscopically confirmed BDs in our sample of 342 low mass candidates above the completeness limit is therefore between 4 and 
 7, and only M-66 does not come from the proper motion selection. 
The calculation for missing BDs is identical to the one above, but with different success rates. 
For the ``$IJ$-pm'' sample we have 4 to 6 BDs over 32 objects, and for the 
photometric candidates 0 to 1 BD over 82. 
The estimated number of unconfirmed BDs is then $6\substack{+10 \\ -3}$.

\subsection{Spectral type distribution}
\label{S_sptdistr}

\begin{figure*}
\centering
\resizebox{14cm}{!}{\includegraphics{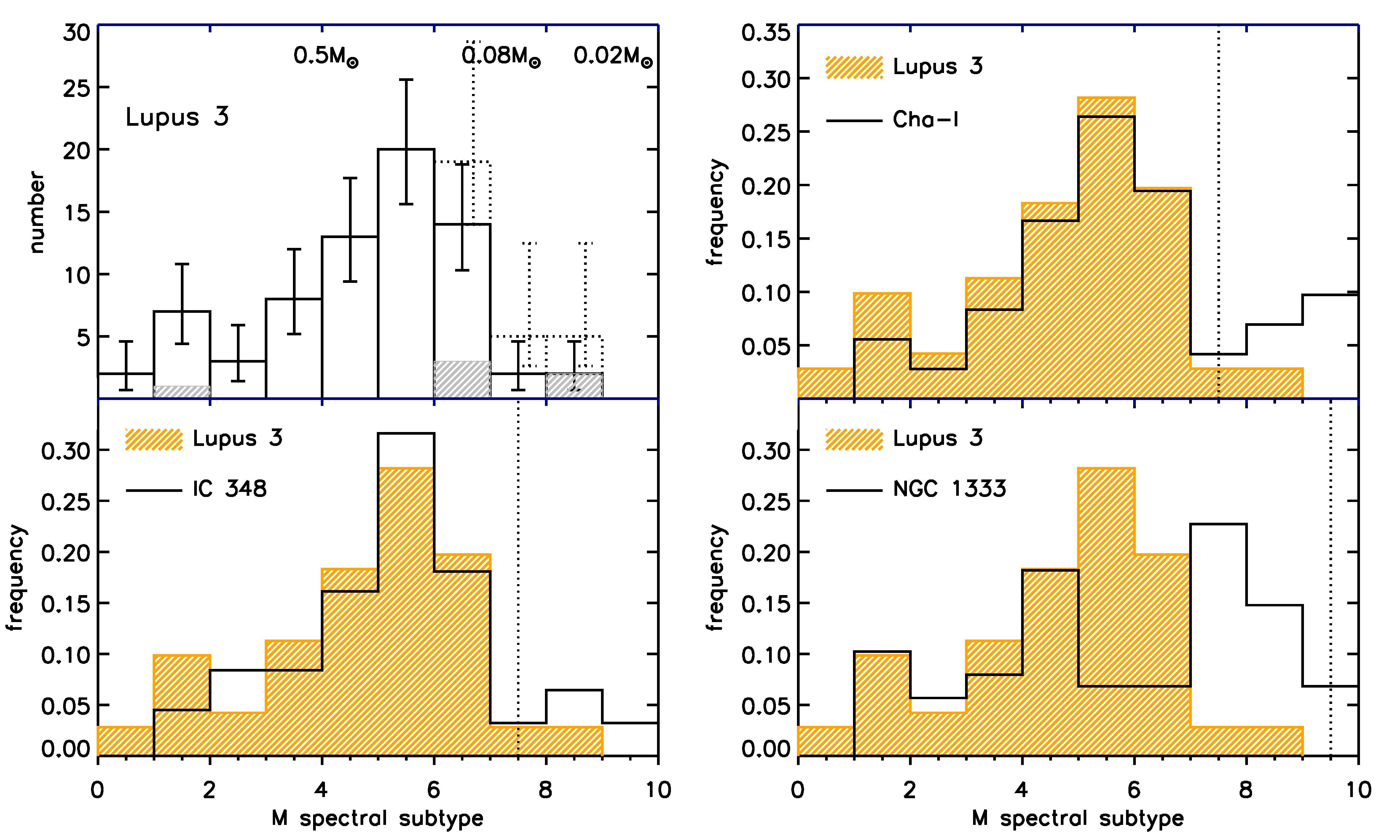}}
\caption{{\it Top left:} Distribution of spectral types for the VLM population of Lupus$\,$3, featuring the M-type sources from Table~\ref{T_census} within the field covered by our survey. Shaded histogram shows the SONYC contribution to the census. Dotted bins take into account the estimate of the objects probably missing in our survey (for clarity, the associated error-bars are displaced to the right of the uncertainty in the number of confirmed sources). The rough mass limits according to BT-Settl and AMES-Dusty isochrones are shown on top of the plot. {\it Top right and two bottom panels:} Comparison between the spectral type distribution in Lupus$\,$3 and those in Cha-I, IC~348 \citep{luhman07} and NGC~1333 \citep{scholz12a}.
}
\label{spt_distr}
\end{figure*}

In Figure~\ref{spt_distr} we show the distribution of spectral types for the spectroscopically
confirmed M-type population in Lupus$\,$3, found within the field covered by our survey (top left panel; solid lines). The solid error bars are Poissonian
confidence intervals calculated with the method described in \citet{gehrels86}, and the dashed histogram shows the contribution of our survey to the census. 
The dotted histogram includes the estimate on the number of missing objects estimated in Section~\ref{missing}, assuming that 6 of the objects are BDs (i.e. found somewhere in the last two bins, and equally distributed for the purpose of this plot), and the remaining 5 are probably low-mass stars, which due to the saturation limit of our survey around 0.06$\,$\solm, 
should fall in the bin above. 
%We remind the reader that the error bars on these estimates are quite large (see Section~\ref{missing}), and were not
%plotted for clarity.
The dotted-line errorbars combine in quadrature the CI from the binomial distribution, with the errorbars shown with solid lines. 

We also show a comparison
with the results from other star forming regions: Cha-I (\citealt{luhman07}; top right), 
IC~348 (\citealt{luhman07}; bottom left), and NGC~1333 (\citealt{scholz12b}; lower right).
In the stellar regime, the distribution resembles those seen
in other star forming regions, with the peak around M5-M6, and 
an increase in the number of objects between M0-M1 and M5 by a factor of 2-3. 
Cha-I and IC~348 show a drop in the number of 
objects in the substellar regime with respect to stars, and this is at a similar level seen also in Lupus$\,$3.
Although this sharp drop at spectral types M7 and later can partially be explained by the 
incompleteness of our survey at lowest masses, it is likely to be a real feature of the IMF considering the 
analysis of the missing objects in our survey presented in the previous section.

In NGC~1333 \citep{scholz12b}, we see a second peak around M7, which might be due to a gap in the completeness between
two different spectroscopic samples that were used in the stellar and substellar regimes, causing us to miss the objects around M6. 
The double peak is also seen in the distribution of absolute J-magnitudes, but this distribution is consistent, within statistical errors, with a single broad peak. On the substellar side, the SONYC survey in NGC~1333 is deeper than the surveys in Cha-I, IC~348 and Lupus$\,$3, and the completeness limit is in fact outside the $x$-axis range in Figure~\ref{spt_distr}. From the statistical analysis in the previous section, given the relatively large errors, we can say that the SpT distribution down to M9 is consistent with that of NGC~1333. 
%In Section~\ref{missing}
% we estimated that the number of missing BDs in our survey is 3-7, down to the completeness limit of 
%0.009 - 0.02 \solm\ for Av=0-5. 
%These missing objects should therefore be somewhere in the two lowest-mass bins of the histogram, or possibly in the one below.
%If we add these objects to the M7-M9 bin, the number of objects between M5 and M7-M9 decreases roughly by a factor of 2-3, 
%in line with findings in other clusters.

\subsection{Star/BD ratio}
\label{SBD}

In this section we use the results of our survey, combined with previous works to get some constraints on the IMF in Lupus$\,$3, extending
down to the completeness limit of our $iJ$ catalog. In the following, we will take into account the area of our combined
$iJ$ survey (solid line in Figure~\ref{IJCMD}). We note that the number of missing objects calculated in the previous section is partially based on the success rates
of the candidates selected from the proper motions, i.e. from a somewhat smaller area than that of the photometric survey. We, however, choose not to 
apply any correction to account for this because, from Figure~\ref{IJCMD}, it is evident that the majority of the high-priority candidates concentrate in the
cloud core, and very few candidate members are expected to be missed in the surroundings. The surveys of \citet{merin08} and \citet{comeron09} have a 
comparable sky coverage in the area around the main Lupus$\,$3 core, and they also cover the northeastern clouds of lower density, that are also considered to be part of the complex.
This portion of the cloud contains a lower concentration of members (only about 10\% of all the members identified in \citealt{merin08} and \citealt{comeron09}), and thus excluding it should not significantly affect our analysis.

For this calculation, we compile a list of all members and candidate members from the two most extensive surveys found in the literature, \citet{merin08} and \citet{comeron09}, and combine it with the sources identified in this work.    
As for the other works, we note that almost all sources from Table~11 of \citet{comeron08} and \citet{comeron03}, and all the confirmed sources by \citet{allen07}, 
are already included either in the Merin et al. list, or the SONYC list. We also include the estimated number of still undiscovered stellar and substellar members of Lupus$\,$3, 
as determined in the previous section.
%A large fraction of the photometric candidate list of H$_{\alpha}$ emitters by \citet{lopez-marti05} is 
%also already accounted
%Also, there are only a few sources from \citet{lopez-marti05} that might be included in the list, and were not
%previously discarded as member by \citet{comeron09}, but we refrain from doing so because 
%In Table~\ref{T_census} we compiled a list of all spectroscopically confirmed members of Lupus$\,$3 with spectral types M0 or later. From this table, we only take into account the sources within the area covered by SONYC. 
%The studies of \citet{mortier11} and \citet{comeron13} represent a spectroscopic follow-up of the candidates identified in \citet{merin08} and \citet{comeron09}, respectively. 

\citet{mortier11} confirm $\sim83\%$ of the MIR excess sources from \citet{merin08} as probable members of Lupus$\,$3, while the confirmation rate for the \citet{comeron09} sample
is $\sim50\%$ \citep{comeron13}. These factors are taken into account when estimating the total number of probable members of the two mentioned studies. 
Another important correction factor that has to be taken into account is the disk fraction, which,
according to \citet{merin08}, is 70-80\% for Lupus. Therefore, the number of objects from this work is additionally multiplied by a factor of 1.25.
%and virtually all the objects identified in that work are stars. 
The sample from \citet{merin08} is complete down to 0.1\solm, and nicely complements the one of \citet{comeron09}, whose method is only sensitive to the objects cooler than 3400$\,$K. The 3$\sigma$ 
detection limit of \citet{comeron09} is at $I_C = 22.2$, but the completeness limits are not specified. However, given the depth of their observations, 
it is probably safe to assume that they are complete in the range of masses between 0.1\solm\ of \citet{merin08}, and the upper limit in mass set by saturation in our optical data ($\sim$0.06 \solm). 

%Since a significant number of the objects from \citet{merin08} lack estimates of spectral types or $T_{\mathrm{eff}}$, we decide not to assess
%numbers of stars and BDs based on spectral types and/or $T_{\mathrm{eff}}$, as we did in the majority of the SONYC papers to date. Here we rather take the 
To assess the numbers of stars and BDs, we use the approach described in \citet{scholz13}.
In short, by comparing the multi-band photometry 
with the predictions of the evolutionary models (BT-Settl in this case), we derive best-fit mass
and A$_V$ for each object, for the assumed distance of 200pc, age of 1 Myr, and the extinction law from \citet{cardelli89} with R$_V=4$. 
We note that the several underluminous objects identified here and in previous studies are erroneously identified by this procedure as BDs. 
However, these are clearly stars, and therefore are counted in this higher-mass bin.

%Merin, in our area: 41 (>0.085) stars, 22 BDs (<0.065). Multiplied by 0.83*1.25 -> 43 stars and 23 BDs
%5 objects with masees between 0.065 and 0.085 (border) -> 5
% => stars 43-48, BDs 23 - 28

%Comeron: 49 stars, 4 BDs. multiplied by 0.5 -> 25 stars, 2 BDs
%5 objects with masees between 0.065 and 0.085 (border) -> 2.5 ~ 3
% => stars 25-28, BDs 2 - 5

%LM05: only a few sources in the area, and with good mass estimate. One of them (L2) rejected as member by Comeron (2011). -> neglect
%Allen: 1 in SONYC, the rest all counted in Merin.
%comeron03: 1 in SONYC, 1 in Merin, 2 stars.
%comeron08: almost all strs from Table11 are already in Merin, except 2 (K-type)
%sonyc: 3 BDS, 4 stars + missing 4-8 stars and 3-7 BDs

%stars: (43-48) + (25-28) + 2 + 2 + 4 + (4-8) = 80 - 92
%BDs:   (23-28) + (2-5) + (3-7) = 28 - 40
%ratio=2.0-3.3

All the objects with estimated masses below 0.065$\,$M$_{\odot}$ are counted as BDs, and all those above 0.085$\,$M$_{\odot}$ as stars. The remaining
objects at the border of the substellar regime are once included in the higher mass bin, and then in the lower mass bin. The calculated number of stars is then 80 - 92, and
the number of BDs is estimated to be 28 - 40. 
Finally, we obtain the star/BD ratio between 2.0 and 3.3, in line with other star-forming regions, 
which typically show a span of values between 2 and 6 (Scholz et al., 2013, in press). 
Clearly, there is a number of uncertainties involved in this calculation due to possible incompleteness at the overlap of the different studies, uncertainties in age and distance,
as well as the choice of the isochrones used to derive masses. It is nevertheless an useful exercise at least for a first order comparison with other works.

\section{Summary and conclusions}
\label{summary}

In this work, we have presented deep optical and near-infrared images of the 1.4$\,$deg$^2$ area surrounding the two brightest members HR5999/6000 of the 
Lupus$\,$3 star forming region. From the optical+NIR photometry we selected 409 candidate VLM cluster members. Proper motion analysis, based on two epochs of
imaging separated by 11-12 years, helped to narrow down the candidate selection to 59 high priority candidates. To confirm the membership of the 
selected candidates, we performed spectroscopic follow-up using VIMOS/VLT, in which we collected 123 spectra from the photometric selection box, including 32
from the high priority list. We confirm 7 candidates as probable members of Lupus$\,$3, among which 4 are later than M6.0 and with $T_{\mathrm{eff}}\leq 3000\,$K, 
i.e. are probably substellar in nature.
Two of the sources identified as probable members of Lupus$\,$3 appear underluminous for their spectral class, similar to some previously known members exhibiting similar
emission line spectra with strong H$_{\alpha}$ and several forbidden lines associated with active accretion.  

We derive a relation between the spectral types (from comparison to low-gravity objects in young regions) and $T_{\mathrm{eff}}$ (from BT-Settl models): 
$T_{\mathrm{eff}}$ = $4083 - 166$ $\times$ SpT, where SpT refers to
the M spectral subtype between 1 and 9. The rms-error of the fit of 61 K.
We derive a star-to-BD ratio of 2.0 - 3.3, consistent with the values observed in other star forming regions. 

Combining our results with previous work on Lupus$\,$3, we compile a table containing all spectroscopically confirmed low-mass objects with spectral type M0 or later, and
show that the distribution of spectral types is in line with what is observed in other young star forming regions.

%\begin{figure*}
%\centering
%\resizebox{16cm}{!}{\includegraphics{underluminous.jpg}}
%\caption{Best fit spectra of young dwarfs (red), giants (blue) and field dwarfs (green) for the (possibly) underluminous objects. {\bf Please comment! SONYC-Lup3-1, -6 , -8, -11, and -12
%show H$_{\alpha}$ emission. A$_V$ is in all cases consistent with them being in Lupus. 
%Pay attention to the 8200\AA Na II feature, to discriminate between spectral classes. Would you discard some of these objects as members? 
%Should we show a plot similar to this one?
%}
%}
%\label{under}
%\end{figure*}

\appendix
\section{Model fitting plots}
\begin{figure*}
\centering
%\sidecaption
\resizebox{16cm}{!}{\includegraphics{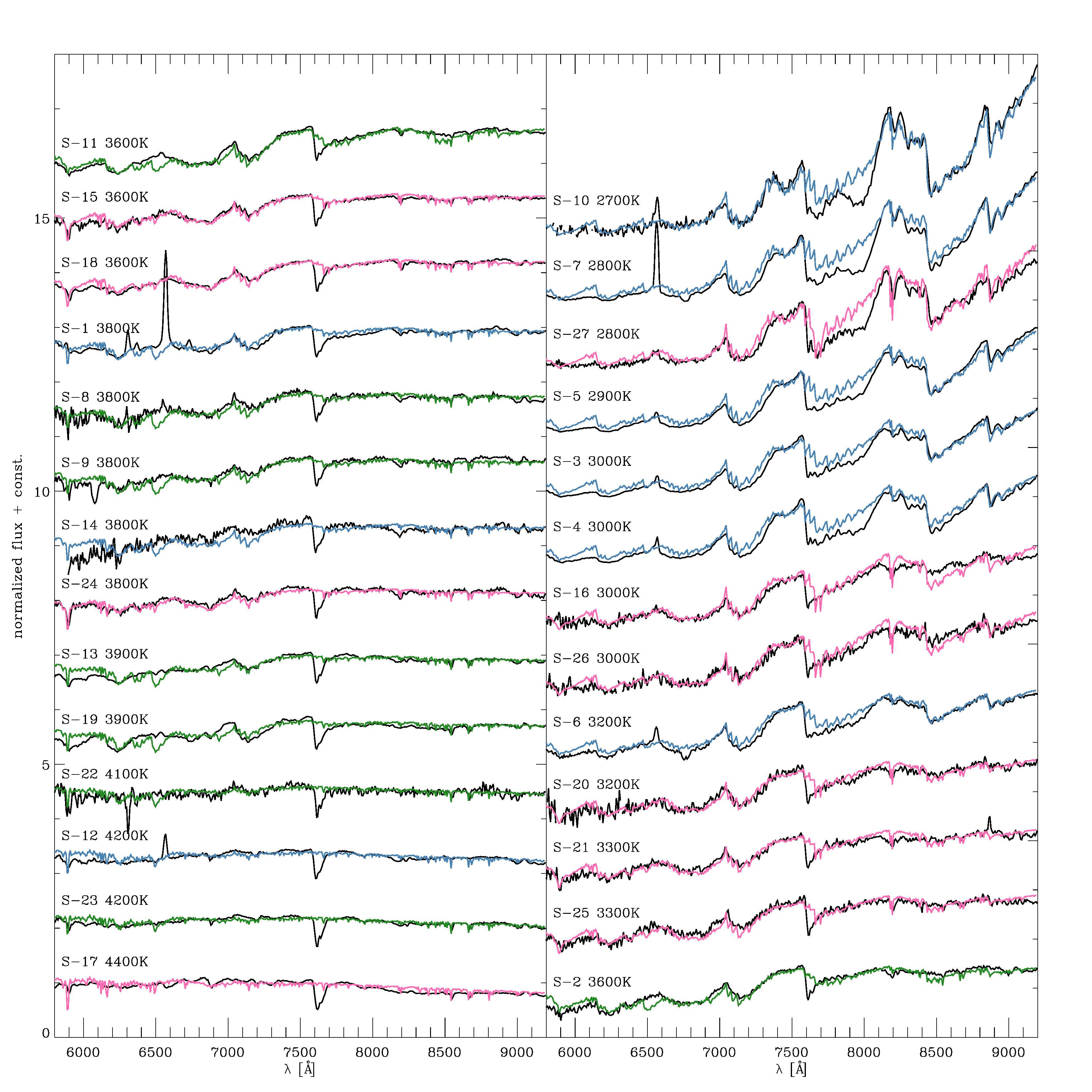}}
\caption{Spectra of the objects in Table~\ref{T_spec} (black), and the best-fit BT-Settl model (\citealt{allard11}). The model spectra
are color-coded according on the log$\,(g)$ used in fitting: green assume log$\,(g)=3.0$, blue log$\,(g)=3.5$, and pink log$\,(g)=5.0$. The spectral resolution of the models has been reduced 
by boxcar smoothing, in order to match the resolution of our spectra. Each spectrum has been corrected
for extinction, but not for the atmospheric absorption. 
}
\label{memb_spec_BT}
\end{figure*}

\begin{figure*}
\centering
%\sidecaption
\resizebox{16cm}{!}{\includegraphics{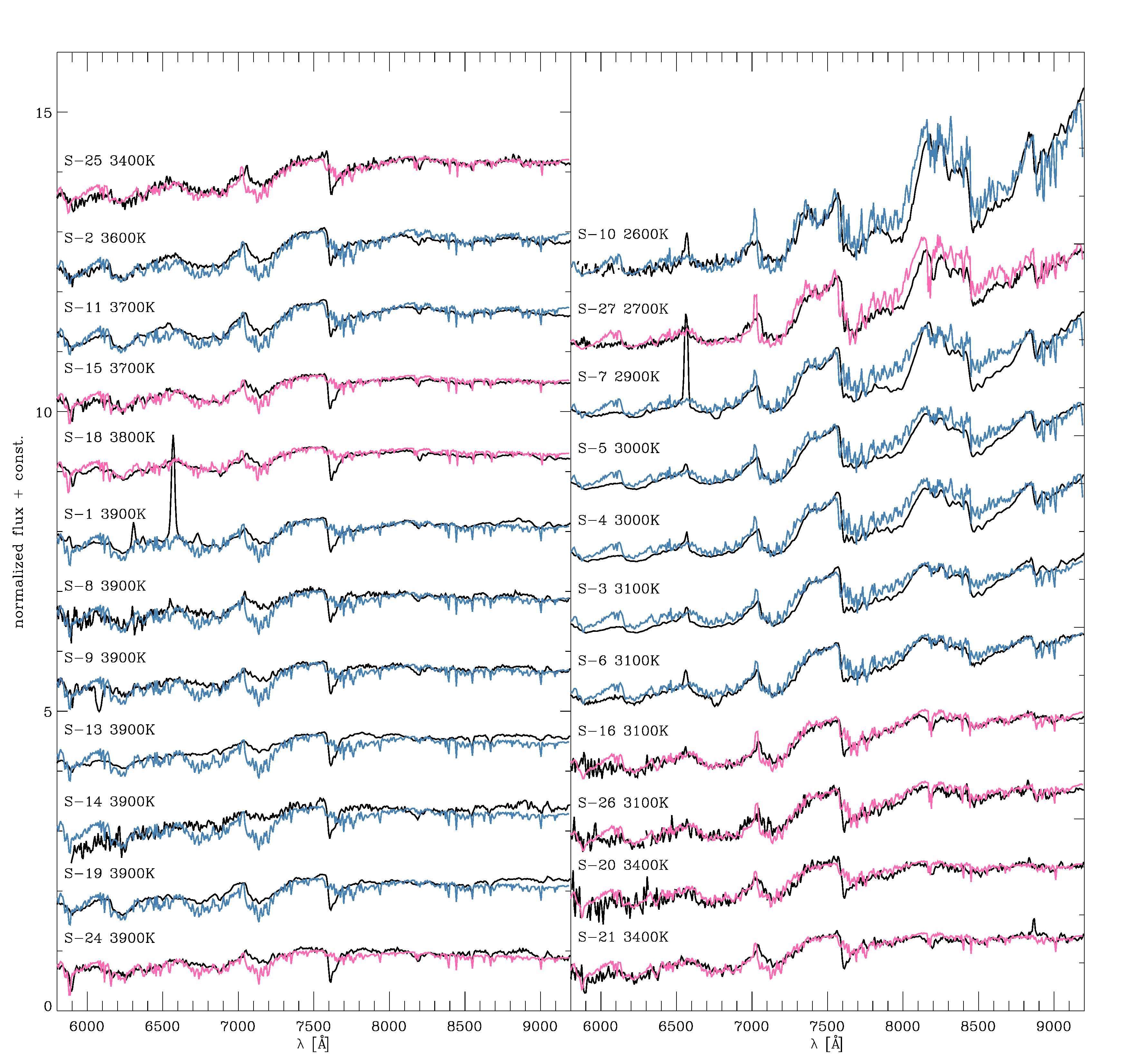}}
\caption{Spectra of the objects in Table~\ref{T_spec} (black), and the best-fit AMES-Dusty model (green; \citealt{allard01}). The model spectra
are color-coded according on the log$\,(g)$ used in fitting, with blue for the low-gravity objects with log$\,(g)=3.5$ or log$\,(g)=4.0$, and pink for the high-gravity objects with log$\,(g)=5.0$. 
The spectral resolution of the models has been reduced 
by boxcar smoothing, in order to match the resolution of our spectra. Each spectrum has been corrected
for extinction, but not for the atmospheric absorption. Note that the AMES-Dusty model do not extend above $T_{\mathrm{eff}}=4000\,$K, therefore the hottest
objects in our sample are not shown in this plot.
}
\label{memb_spec_AD}
\end{figure*}

Figures~\ref{memb_spec_BT} and \ref{memb_spec_AD} show the best-fit models from the BT-Settl and 
AMES-Dusty series to the spectra of the objects from Table~\ref{T_spec}. The results are summarized in Table~\ref{T_models}. For
more details please refer to Section~\ref{model_fit}.

\acknowledgements{We thank Kevin Luhman and Amelia Bayo for sharing the spectra of young low-mass objects. 
This work was co-funded under the Marie Curie Actions of the European
Commission (FP7KCOFUND). AS acknowledges financial support through the grant
10/RFP/AST2780 from the Science Foundation Ireland.
B.L.M. was supported by the Spanish \emph{Plan Nacional de Astronom\'{i}a
y Astrof\'{i}sica} through project AYA 2011-30147-C03-03. 
Support for this work also came from grants to RJ from the Natural Sciences and Engineering Research Council of Canada.
}

\begin{deluxetable}{lllccccl}
\tabletypesize{\scriptsize}
%\rotate
\tablecaption{Spectroscopically confirmed very low-mass members of Lupus$\,$3}
\tablewidth{0pt}
\tablehead{\colhead{ID} & 
	   \colhead{$\alpha$(J2000)} & 
	   \colhead{$\delta$(J2000)} & 
	   \colhead{$J$} & 
	   \colhead{$K$} & 
	   \colhead{SpT} & 
	   \colhead{$T_{\mathrm{eff}}$}	&
	   \colhead{other names} \\
&	      &		        &(mag) & (mag) & & (K) &}
\tablecolumns{8}
\startdata
SONYC-Lup3-1  & 16 07 08.55 &  -39 14 07.8 & 14.6 & 11.5 & M1.75$^a$     & 3850$^a$, 4400$^b$ &\nodata \\ 	
SONYC-Lup3-3  & 16 07 55.18 &  -39 06 04.0 & 13.5 & 11.4 & M$\,$5.75$^a$, M6.5$^c$ & 3050$^a$, 2900$^b$, 2950$^c$ & \nodata\\ 
SONYC-Lup3-4  & 16 08 04.76 &  -39 04 49.6 & 13.9 & 12.4 & M$\,$6.50$^a$, M7$^c$ & 3000$^a$, 2700$^b$, 2900$^c$ & \nodata\\
SONYC-Lup3-5  & 16 08 16.03 &  -39 03 04.5 & 12.5 & 11.1 & M$\,$6.25$^a$, M7.5$^d$,M5.5$^g$ & 2950$^a$, 2800$^b$ & Par-Lup3-1   \\ 
SONYC-Lup3-7  & 16 08 59.53 &  -38 56 27.8 & 13.9 & 12.7 & M$\,$8.00$^{a,e}$, M8.5$^h$  & 2850$^a$, 2600$^{b,h}$ & \nodata\\
SONYC-Lup3-10 & 16 09 13.43 &  -38 58 04.9 & 16.1 & 14.9 & M$\,$8.75$^a$ & 2650$^a$	& \nodata      \\
SONYC-Lup3-12 & 16 11 18.47 &  -39 02 58.2 & 15.1 & 13.3 & $<$M$\,$1$^a$ & 4200$^a$& \nodata	      \\
EX Lup & 16 03 05.5 & -40 18 26 & 9.7 & 8.5 & M0$^f$ & \nodata& HD 325367, Th$\,$14\\ 	
Sz-88 & 16 07 00.61 & -39 02 19.9 & 9.9  & 8.6  &  M1$^f$, M0$^h$ & 3850$^h$ & Th$\,$15 \\
Sz-91 & 16 07 11.57 & -39 03 47.2 & 11.1 & 9.8  &  M0.5$^f$, M1$^h$ & 3705$^h$ & Th$\,$20 \\
Sz-94 & 16 07 49.68 & -39 04 28.6 & 11.4 & 10.6 &  M4$^f$, M4.5$^g$ &\nodata & \nodata \\
Sz-95 & 16 07 52.29 & -38 58 05.6 & 11.0 & 10.0 &  M1.5$^f$ & \nodata& \nodata \\
Sz-96 & 16 08 12.48 & -39 08 33.0 & 10.1 & 9.0  &  M1.5$^f$ ,M2$^g$  & 3560$^g$ & \nodata\\
Sz-97 & 16 08 21.84 & -39 04 21.4 & 11.2 & 10.2	&  M3$^f$,M4.5$^g$, M4$^h$   & 3198$^g$, 3270$^h$ &  Th$\,$24 \\
Sz-98 & 16 08 22.44 & -39 04 46.7 & 9.5	 & 8.0  &  M0$^f$ & \nodata&\nodata  \\
Sz-99 & 16 08 24.00 & -39 05 49.2 & 11.9 & 10.7 &  M3.5$^f$,M4$^{g,h}$  & 3270$^{g,h}$ & Th$\,$25\\
Sz-100 & 16 08 25.76 & -39 06 01.2 & 11.0 & 9.9 &  M5$^{d,f}$, M4.5$^g$, M5.5$^h$  & 3198$^g$, 3057$^h$ & Th$\,$26 \\ 
Sz-101 & 16 08 28.37 & -39 05 32.3 & 10.4 & 9.4 &  M4.5$^g$,M4$^f$ & 3198$^g$ & Th$\,$27 \\
Sz-103 & 16 08 30.24 & -39 06 10.8 & 11.4 & 10.2 & M4$^{f,h}$, M4.5$^g$   & 3198$^g$, 3270$^h$ & Th$\,$29\\
Sz-104 & 16 08 30.72 & -39 05 48.5 & 11.7 & 10.7 & M5$^{f,h}$,M5.5$^g$   & 3058$^g$, 3125$^h$ & Th$\,$30\\
Sz-105 & 16 08 37.03 & -40 16 20.8 & 9.0  &  7.6 & M4$^f$ &\nodata & Th$\,$31 \\ 
Sz-106 & 16 08 39.76 & -39 06 25.3 & 11.7 & 10.1 & M0$^f$,M2.5$^d$, M0.5$^h$  & 3777$^h$ & \nodata \\
Sz-107 & 16 08 41.76 & -39 01 36.8 & 11.2 & 10.3 & M5.5$^f$, M6.5$^g$   & 2935$^g$ & \nodata \\
Sz-108A	& 16 08 42.74 & -39 06 18.4 & 9.8 &  8.8 &  M3$^d$   & \nodata&  \nodata\\
Sz-108B	& 16 08 42.87 & -39 06 14.6 &\nodata &\nodata &  M6$^d$, M5.5$^g$   & 3058$^g$ & \nodata\\   
Sz-109 & 16 08 48.16 & -39 04 19.3 & 11.4 & 10.5 &  M5.5$^f$, M6.5$^d$ &\nodata & \nodata \\
Sz-110 & 16 08 51.57 & -39 03 17.7 & 11.0 & 9.7  &  M4.5$^d$, M3$^g$, M2$^f$,M4$^h$  & 3415$^g$, 3270$^h$  & Th$\,$32\\  
Sz-111 & 16 08 54.74 & -39 37 43.6 & 10.6 & 9.5  & M1.5$^f$, M1$^h$  & 3705$^h$ & Th$\,$33\\
Sz-112 & 16 08 55.53 & -39 02 34.0 & 11.0 & 10.0 &  M4$^f$,M6$^d$, M5$^h$  & 3125$^h$  & \nodata \\
Sz-113 & 16 08 57.80 & -39 02 22.8 & 12.5 & 11.3 &  M4$^f$, M6$^d$, M1.5$^g$, M4.5$^h$    & 3633$^g$, 3197$^h$  & Th$\,$34, Lup 609s\\ %84
Sz-114 & 16 09 01.85 & -39 05 12.4 & 10.4 & 9.3  &  M5.5$^d$, M4$^{g,f}$, M4.8$^h$  & 3270$^g$, 3175$^h$  & Th$\,$35\\
Sz-115 & 16 09 06.23 & -39 08 51.9 & 11.3 & 10.4 &  M4$^f$, M4.5$^h$  & 3197$^h$ & \nodata\\
Sz-116 & 16 09 42.61 & -39 19 41.5 & 10.5 & 9.5  & M1.5$^f$ & \nodata& Th$\,$36\\
Sz-117 & 16 09 44.34 & -39 13 30.4 & 10.7 & 9.4	 & M2$^f$& \nodata& Th$\,$37\\
Sz-119 & 16 09 57.07 & -38 59 47.6 & 10.4 & 9.4  & M4$^f$& \nodata& Th$\,$38\\
Sz-121 & 16 10 12.21 & -39 21 18.6 & 10.0 & 9.0	 & M3$^f$& \nodata& Th$\,$40\\
Sz-122 & 16 10 16.44 & -39 08 05.4 & 10.9 & 9.9  & M2$^f$& \nodata& Th$\,$41\\
Sz-123 & 16 10 51.49 & -38 53 14.1 & 11.1 & 9.8  & M1$^{f,h}$& 3705$^h$ & Th$\,$42\\
%Par-Lup3-1 &	16 08 15.9 & -39 03 07 & & &  M7.5$^d$ & & \\
Par-Lup3-2 &	16 08 35.78 & -39 03 47.9 & 11.2 & 10.3	 &  M6$^d$, M5$^g$   & \nodata& \nodata \\ 
Par-Lup3-3 &	16 08 49.40 & -39 05 39.3 & 11.4 & 9.5   &  M4.5$^d$, M4$^h$  & 3270$^h$  & \nodata\\
Par-Lup3-4 &	16 08 51.44 & -39 05 30.5 & 15.5 & 13.3  &  M5$^d$, M4.5$^h$    & 3197$^h$ & \nodata\\
Lup 713s   &    16 07 37.73 & -39 21 38.8 & 13.2 & 12.1  &  M5.75$^e$, M5.5$^h$  & 3057$^h$ & \nodata \\
Lup 604s   &    16 08 00.17 & -39 02 59.5 & 12.1 & 11.1	 &  M5.25$^e$, M5.5$^{g,h}$ & 3058$^{g,h}$ & \nodata\\ 
J16081497-3857145	   & 	16 08 14.97 & -38 57 14.5 & 15.2 & 13.1  &  M4.75$^e$  & \nodata& \nodata\\
Lup 706    &    16 08 37.33 & -39 23 10.9 & 15.2 & 13.8  &  M7.75$^e$, M7.5$^h$  & 2795$^h$ & \nodata\\	
J16085373-3914367	  &	16 08 53.73 & -39 14 36.7 & 15.0 & 12.5	 &  M5.5$^e$  & \nodata& \nodata\\
%	   & 	16 08 59.53 & -38 56 27.5 & & &  M8$^e$  & & \\
%    M-3    &    16 08 06.24  & -39 12 22.3  & 10.0 & 7.7   &   M6.5$^g$   & 2935$^g$ & \\
    M-7    &    16 08 55.20  & -38 48 48.2  & 13.0 & 12.0  &     M3$^g$   & 3415$^g$ & \nodata\\
%   16    &    16 08 00.24  & -39 02 58.9  & & &   M5.5$^g$   & & \\
   M-17    &    16 08 28.08  & -39 13 09.8  & 13.7 & 12.4 &   M5.5$^g$   & 3058$^g$ & Lup 607 \\
   M-18    &    16 09 08.40  & -39 03 42.8  & 12.2 & 11.4  &   M5.5$^g$   & \nodata& Lup 608s \\
   M-19    &    16 08 48.24  & -39 09 19.4  & 12.9 & 12.0  &     M6$^g$   & \nodata& Lup 617 \\
   M-20    &    16 09 49.92  & -38 49 02.6  & 15.4 & 14.4  &   M3.5$^g$   &\nodata & Lup 650 \\
   M-23    &    16 09 17.04  & -39 27 09.7  & 13.5 & 12.7  &   M4.5$^g$   &\nodata & Lup 710\\
   M-24    &    16 07 58.80  & -39 24 34.9  & 12.6 & 11.6  &     M6$^g$   &\nodata & Lup 714 \\
   M-25    &    16 11 51.12  & -38 51 05.0  & 13.3 & 12.3  &   M5.5$^g$   & \nodata& Lup 802s\\
 %  26    &    16 09 54.48  & -39 12 03.2  & & &     K0$^g$   & & \\
   M-27    &    16 09 56.40  & -38 59 51.0  & 13.0 & 12.0  &     M6$^{g,h}$   & 2990$^{g,h}$ & Lup 818s\\
 %  47    &    16 08 16.08  & -39 03 04.0  & & &   M5.5$^g$   & & \\
%   48    &    16 08 35.76  & -39 03 47.5  & & &     M5$^g$   & & \\
   M-51    &    16 11 59.76  & -38 23 38.4  & 12.2 & 11.2  &     M6$^g$, M5$^h$    & 2990$^g$, 3125$^h$  & SST-Lup3-1\\
   M-58    &    16 07 03.84  & -39 11 11.4  & 14.7 & 13.1  &   M5.5$^g$   & 3058$^g$ & \nodata\\
%   M-60    &    16 07 55.20  & -39 07 17.4  & 10.3 &  7.6  &     M9$^g$   & 2400$^g$ & \\
%   M-61    &    16 08 03.12  & -38 52 29.6  & 9.5	 & 8.1   &   M5.5$^g$   & 3058$^g$ & \\
   M-62    &    16 09 01.44  & -39 25 12.0  & 11.6 & 10.3  &   M3.5$^g$, M4$^h$  & 3343$^g$, 3270$^h$  & \nodata\\
%   M-63    &    16 09 34.08  & -39 13 42.2  & 8.7	 & 6.7   &   M7.5$^g$   & 2795$^g$ & \\
%   M-64    &    16 10 00.00  & -38 54 00.4  & 9.6	 & 7.7   &   M6.5$^g$   & 2935$^g$ & \\
   M-65    &    16 10 12.96  & -38 46 16.3  & 15.9 & 13.9  &     M4$^g$   & 3270$^g$ & \nodata\\
   M-66    &    16 10 19.92  & -38 36 06.5  & 13.3 & 12.3  &     M6$^g$   & 2990$^g$ & \nodata\\
   M-67    &    16 10 29.52  & -39 22 14.5  & 11.9 & 10.9  &   M5.5$^g$   & 3058$^g$ & \nodata\\
%   M-68    &    16 10 34.56  & -38 14 50.3  & 8.0	 & 6.3   &   M6.5$^g$   & 2935$^g$ & \\
   M-69    &    16 11 31.92  &-38 11 10.0   & 12.3 & 10.7  &   M2.5$^g$   & 3488$^g$ & \nodata\\
   M-70    &    16 11 44.88  & -38 32 44.9  & 12.4 & 11.5  &   M6.5$^g$   & 2935$^g$ & \nodata\\
%  71    &    16 11 48.72  & -38 17 57.8  & & &     K2$^g$   & & \\
%   M-74    &    16 12 19.68  & -38 37 41.9  & 8.7  & 7.2  &     M8$^g$   & 2710$^g$ & \\
%   M-75    &    16 12 51.60  & -38 42 15.8  & 8.6	 & 7.0  &   M8.5$^g$   & 2555$^g$ & \\
%  76    &    16 08 25.68  & -39 06 01.1  & & &   M4.5$^g$   & & \\
%   77    &    16 08 28.32  & -39 05 31.9  & & &   M4.5$^g$   & & \\
%   78    &    16 08 29.76  & -39 03 10.8  & & &     K2$^g$   & & \\
%   79    &    16 08 30.24  & -39 06 10.8  & & &   M4$^f$, M4.5$^g$   & & Sz 103\\
%   80    &    16 08 30.72  & -39 05 48.5  & & &   M5$^f$,M5.5$^g$   & & Sz 104 \\
%   81    &    16 08 41.76  & -39 01 36.8  & & &   M5.5$^f$, M6.5$^g$   & & Sz 107 \\
%   82    &    16 08 42.96  & -39 06 14.4  & & &   M5.5$^g$   & & \\
%   83    &    16 08 51.60  & -39 03 17.6  & & &     M3$^g$   & & \\
%   84    &    16 08 57.84  & -39 02 22.6  & & &   M1.5$^g$   & & \\
%   85    &    16 09 01.92  & -39 05 12.1  & & &     M4$^g$   & & \\
%   86    &    16 09 48.72  & -39 11 16.8  & & &     K7$^g$   & & \\
%   89    &    16 07 49.68  & -39 04 28.6  & & &  M4$^f$, M4.5$^g$   & & Sz 94\\
%   90    &    16 08 12.48  & -39 08 33.0  & & &   M1.5$^f$ ,M2$^g$   & & Sz 96\\
%   91    &    16 08 21.84  & -39 04 21.4  & & &   M3$^f$,M4.5$^g$   & & Sz 97, Th$\,$24 \\
%   92    &    16 08 22.56  & -39 04 45.8  & & &     K5$^g$   & & \\
%   93    &    16 08 24.00  & -39 05 49.2  & & &    M3.5$^f$,M4$^g$   & & Sz 99, Th$\,$25\\
J16052862-3846210    & 16 05 28.63  & -38 46 21.1  & 12.6 & 11.8 &     M6$^c$ &  3000$^c$ &  \nodata\\
%    & 16 07 55.2  & -39 06 03  & & &   M6.5$^c$ &  2950 &  \\
%    & 16 08 04.8  & -39 04 49  & & &     M7$^c$ &  2900  & \\
J16083304-3852224    & 16 08 33.05  & -38 52 22.4  & 12.9 & 11.9 &   M7.5$^c$ &  2800$^c$  & \nodata\\
J16083547-3900358    & 16 08 35.48  & -39 00 35.8  & 11.5 & 10.6 &   M6.5$^c$ &  2950$^c$  & \nodata\\
 %   & 16 08 36.2  & -39 23 02  & & &     K5$^c$ &  4400 &  \\
J16083974-3929228   & 16 08 39.75  & -39 29 22.9  & 11.7 & 10.9 &     M5$^c$ &  3100$^c$ & \nodata \\
J16085575-3826330   & 16 08 55.75  & -38 26 33.1  & 13.0 & 12.1 &   M5.5$^c$ &  3050$^c$ & \nodata \\
J16090452-3921125    & 16 09 04.52  & -39 21 12.5  & 10.7 &  9.8 &     M5$^c$ &  3100$^c$ & \nodata \\
J16091570-3851396    & 16 09 15.71  & -38 51 39.7  & 12.0 & 11.2 &     M5$^c$ &  3100$^c$ & \nodata \\
J16092279-3855506   & 16 09 22.80   & -38 55 50.6  & 10.8 & 9.9 &     M5$^c$ &  3100$^c$ &  J160922.8-385550A$^c$\\
J16092320-3855547   & 16 09 23.20   & -38 55 54.8  & 10.8 & 9.9 &     M4$^c$ &  3250$^c$ &  J160922.8-385550B$^c$\\
J16092800-3848538    & 16 09 28.01  & -38 48 53.8  & 11.9 & 11.0	 &   M5.5$^c$ &  3050$^c$ &  \nodata\\
J16101386-3759589    & 16 10 13.87  & -37 59 58.9  & 12.2 & 11.2  &   M6$^c$ &  3000$^c$ & \nodata \\
J16103062-3831517    & 16 10 30.62  & -38 31 51.7  & 13.3 & 12.4  &     M6$^c$ &  3000$^c$ &  \nodata\\
J16103323-3830234    & 16 10 33.23  & -38 30 23.5  & 10.8 & 9.8   &   M4.5$^c$ &  3200$^c$ & \nodata \\
J16104192-3823046    & 16 10 41.92  & -38 23 04.7  & 10.9 & 10.0  &     M5$^c$ &  3100$^c$ & \nodata \\
J16105899-3914514    & 16 10 58.99  & -39 14 51.4  & 12.1 & 11.3  &     M5$^c$ &  3100$^c$ & \nodata \\
J16113801-3841356    & 16 11 38.02  & -38 41 35.7  & 10.9 & 10.1  &   M2.5$^c$ &  3500$^c$ & \nodata \\
J16120761-3813242    & 16 12 07.61  & -38 13 24.3  & 10.9 & 10.0  &   M4.5$^c$ &  3200$^c$ & \nodata \\
J16121046-3909040    & 16 12 10.46  & -39 09 04.0  & 13.9 & 12.8  &   M6.5$^c$ &  2950$^c$  & \nodata\\
J16122559-3817428    & 16 12 25.58  & -38 17 42.8  & 13.1 & 12.2  &     M6$^c$ &  3000$^c$ & \nodata\\
\enddata
\tablecomments{
Identifiers are from \citet[][Sz]{schwartz77}, \citet[][Par-Lup3]{comeron03}, \citet[][Lup]{lopez-marti05}, \citet[][M]{mortier11}, and 2MASS (J16).}
\tablerefs{
(a) this work, (b) \citet{comeron09}, (c) \citet{comeron13}, (d) \citet{comeron03}, (e) \citet{allen07}, (f) \citet{hughes94}, (g) \citet{mortier11}, (h) \citet{alcala14}.   
}
\label{T_census}
\end{deluxetable}

\end{document}